\documentclass[lettersize,journal]{IEEEtran}
\ifCLASSOPTIONcompsoc
  \usepackage[nocompress]{cite}
\else
  \usepackage{cite}
\fi
\ifCLASSINFOpdf
\else
\fi
\usepackage{amsmath}
\usepackage{amssymb}
\usepackage{amsthm}
\usepackage{amsfonts, xspace,  comment}
\usepackage{algorithmic}
\usepackage{textcomp}
\usepackage{stfloats}
\usepackage{url}
\usepackage{verbatim}

\def\BibTeX{{\rm B\kern-.05em{\sc i\kern-.025em b}\kern-.08em
    T\kern-.1667em\lower.7ex\hbox{E}\kern-.125emX}}
\usepackage{balance}
\usepackage{array, subfigure}
\usepackage{graphicx}
\usepackage{xcolor, soul}
\usepackage[hidelinks]{hyperref}
\usepackage{multirow}
\usepackage{colortbl}
\usepackage{hhline}
\usepackage{booktabs}
\usepackage{enumitem}
\usepackage{float}
\usepackage{caption}
\usepackage{fancyhdr}
\usepackage{orcidlink}

\newcommand{\ours}{{AdvEdge}\xspace}
\newcommand{\oursplus}{{AdvEdge$^{+}$}\xspace}

\DeclareMathOperator*{\argmin}{argmin}

\newcommand{\eg}{{\em e.g.,}\xspace}
\newcommand{\ie}{{\em i.e.,}\xspace}
\newcommand{\etc}{{\em etc.}\xspace}
\newcommand{\BfPara}[1]{{\noindent\bf#1.}\xspace}

\usepackage{tikz}

\newcommand*\cib[1]{\tikz[baseline=(char.base)]{
                            \node[shape=circle,fill=black,text=white,draw,inner sep=0.3pt] (char) {#1};}}

\usepackage{tcolorbox}
\tcbuselibrary{theorems}

\newtcbtheorem[]{observation}{\textbf{Observation}}%
{colback=white,colframe=white!65!black,fonttitle=\bfseries,coltitle=black}{}

\hypersetup{
	colorlinks=true,
	urlcolor=blue!70!black,
	linkcolor=red!70!black,
	citecolor=blue!70!black,
} 

\begin{document}

\title{Interpretations Cannot Be Trusted: Stealthy and Effective Adversarial Perturbations against Interpretable Deep Learning}


\author{Eldor~Abdukhamidov \orcidlink{0000-0001-8530-9477},
        Mohammed~Abuhamad~\orcidlink{0000-0002-3368-6024},
        Simon~S.~Woo~\orcidlink{0000-0002-8983-1542},
        Eric Chan-Tin~\orcidlink{0000-0001-8367-5836},
        and~Tamer~ABUHMED~\orcidlink{0000-0001-9232-4843}
        
        \IEEEcompsocitemizethanks{\IEEEcompsocthanksitem Eldor Abdukhamidov and Tamer ABUHMED are with Department of Computer Science and Engineering, Sungkyunkwan University, Suwon, South Korea.\protect
        (E-mail: abdukhamidov@skku.edu, 
        E-mail: tamer@skku.edu). Simon S. Woo is with Department of Artificial Intelligence and Department of Applied Data Science, Sungkyunkwan University, Suwon, South Korea.\protect
        (E-mail: swoo@g.skku.edu). Mohammed Abuhamad and Eric Chan-Tin are with Department of Computer Science, Loyola University, Chicago, USA.\protect
        (E-mail: mabuhamad@luc.edu,
        E-mail: chantin@cs.luc.edu).\\     }
        \thanks{{$\bullet$ An earlier version of this work has appeared in CSoNET 2021 \cite{Eldor}}}
}


\markboth{preprint}%
{How to Use the IEEEtran \LaTeX \ Templates}

\IEEEtitleabstractindextext{%
\begin{abstract}
Deep learning methods have gained increased attention in various applications due to their outstanding performance. 
For exploring how this high performance relates to the proper use of data artifacts and the accurate problem formulation of a given task, interpretation models have become a crucial component in developing deep learning-based systems. Interpretation models enable the understanding of the inner workings of deep learning models and offer a sense of security in detecting the misuse of artifacts in the input data. 
Similar to prediction models, interpretation models are also susceptible to adversarial inputs. This work introduces two attacks, \ours and \oursplus, that deceive both the target deep learning model and the coupled interpretation model. 
We assess the effectiveness of proposed attacks against two deep learning model architectures coupled with four interpretation models that represent different categories of interpretation models. Our experiments include the attack implementation using various attack frameworks. We also explore the potential countermeasures against such attacks.
Our analysis shows the effectiveness of our attacks in terms of deceiving the deep learning models and their interpreters, and highlights insights to improve and circumvent the attacks. 
\end{abstract}

}

\maketitle
\IEEEdisplaynontitleabstractindextext
\IEEEpeerreviewmaketitle

\section{Introduction}\label{sec:introduction}

\IEEEPARstart{D}{eep} Neural Networks (DNNs) have been increasingly incorporated into a wide range of applications due to their high predictive accuracy in various machine learning tasks, including image classification \cite{he2016deep}, and natural language processing \cite{sutskever2014sequence}. 
Despite these extraordinary achievements, it is not yet entirely evident how the DNN models make certain decisions due to their complex network architecture. Additionally, the susceptibility to adversarial manipulations is another shortcoming of DNN models. For example, adversarial examples, maliciously-crafted inputs, can lead to unexpected model behavior during decision-making.

\begin{figure}[t]
    \centering
    \captionsetup{justification=justified}
    \includegraphics[width=0.45\textwidth]{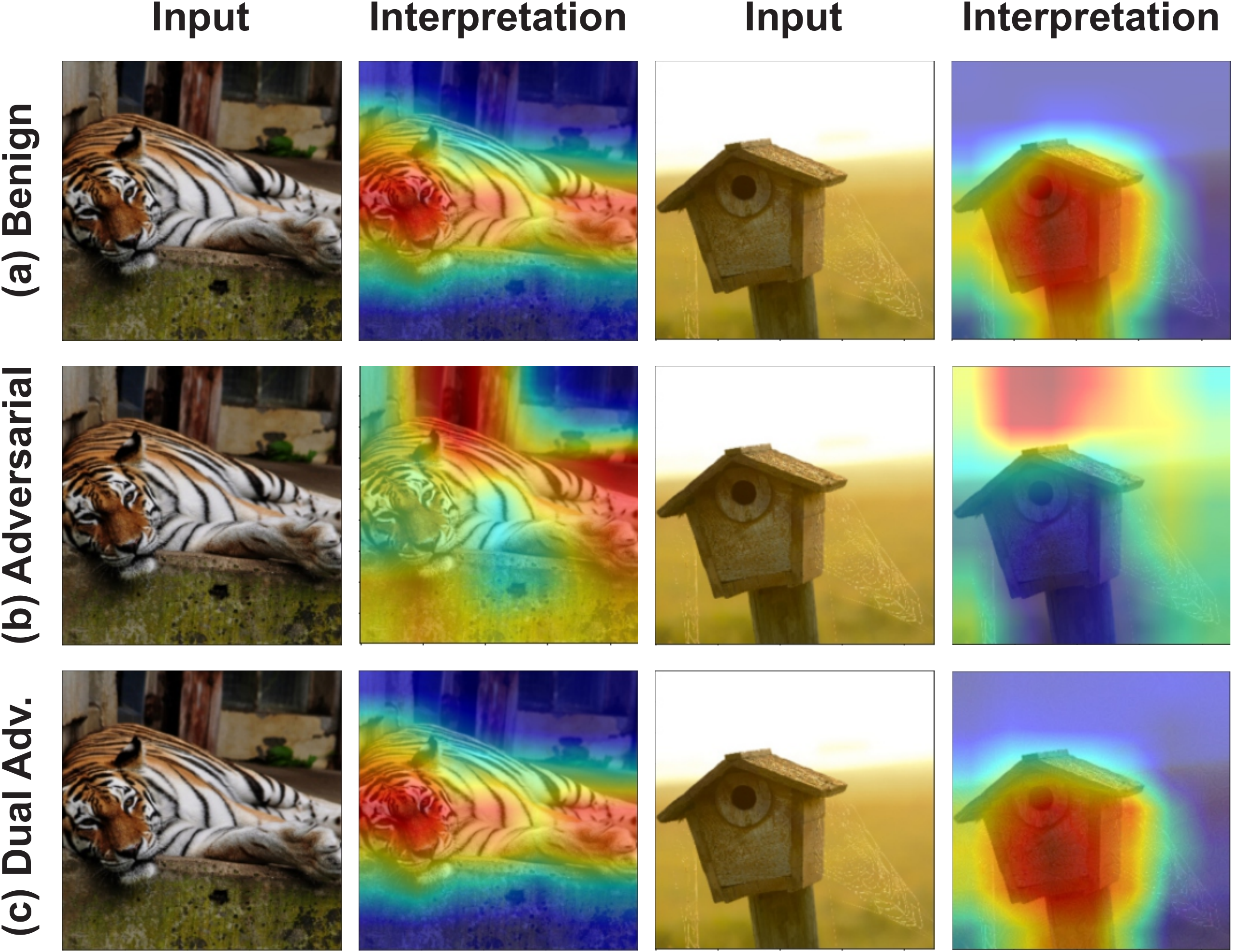}
    \caption{Examples of (a) benign, (b) standard adversarial and (c) dual adversarial images together with their interpretations based on ResNet-50 classifier and CAM interpreter.}
    \label{fig:intro_images}
\end{figure}

Recently, researchers and practitioners have used interpretability as an indispensable tool to describe the behavior of DNN models in an understandable form to humans. Moreover, the high performance of the security-sensitive models is strongly related to interpretability. Several studies have been conducted to improve the security of DNNs by providing interpretability at different levels (\eg overall model \cite{sabour2017dynamic, zhang2018interpretable} and instance levels \cite{dabkowski2017real, fong2017interpretable, simonyan2013deep}). For instance, Figure \ref{fig:intro_images} (a) displays the causal relationship of an attribution map highlighting important regions of an image based on the prediction output. 
Generally, the use of interpretations assists in understanding the internal structure of models to debug them, perform security analysis, and identify adversarial examples. Figure \ref{fig:intro_images} (b) shows successful adversarial examples producing different interpretation maps from the benign interpretation maps.


In the interpretable deep learning system workflow, a classifier (DNN model) and its coupled interpreter (interpretation model) construct interpretable deep learning systems (IDLS). IDLSes offer additional reliability and security in human-assisted decision-making processes, since experts can determine if an interpretation map fits the prediction of the DNN models. Nevertheless, interpretability is also vulnerable to adversarial modification. 
It is both possible and practical to create an adversarial input to fool the target DNN model and its coupled interpretation model simultaneously. 
Figure \ref{fig:intro_images} (c) shows dual adversarial examples resulting from targeting the DNN model and the coupled interpreter. The dual adversarial examples generate interpretation maps indistinguishable from their benign interpretation maps.


In this work, we proposed optimized adversarial attacks that target IDLSes, \ie can fool the prediction model and the coupled interpreter simultaneously. We call these attacks \ours{} and \oursplus{}. In particular, the attacks exploit the image's edge information to enable perturbation to be attached to the edges of the image areas spotlighted by the interpretation model's attribution map. This allows for a far more stealthy adversarial attack since the crafted adversarial inputs are difficult to identify, even with the interpretation and human involvement. Furthermore, the proposed attacks generate effective adversarial examples with minimal perturbation.


\BfPara{Our Contribution} 
In this work, we show that adversarial samples can be used to fool the existing IDLSes. We present two attacks, \ie \ours{} \oursplus{}, that mislead the target DNN and deceive the coupled interpreter using adversarial inputs that have a high attack success rate with less amount of noise. 
We also employ several attack frameworks, \ie PGD, C\&W, and StAdv, for our attacks and evaluate the performance against four interpreters, \ie Grad, CAM, RTS, and MASK.
We evaluate the attacks on the  ImageNet dataset as well as compare them with the existing attack ADV$^2$ \cite{zhang2020interpretable}. 
To present the validity of the attacks, we include additional evaluation metrics in terms of classification, interpretation, and noise generation. In addition, we show that our attacks provide significantly better transferability over ADV$^2$ by testing the attack transferability across different interpreters.  
Finally, we propose two possible countermeasures against our attacks.

Our contributions are summarized as follows:
\begin{itemize}[leftmargin=2em]
\item We propose two novel attacks against IDLSes. Our proposed attacks,\ours and \oursplus, utilize edge information to enhance interpretation-derived attacks. We evaluate the effectiveness of attacks when various DNN models (\eg ResNet and DenseNet) or interpretation models (\eg Grad, CAM, RTS, and MASK) are involved. 

\item We evaluated our attacks on three frameworks (\ie PGD, C\&W, and StAdv) to show that the attacks can be flexibly constructed using various frameworks.

\item 
Experimental findings indicate that \ours{} and \oursplus{} are equally effective as ADV$^2$ in terms of misclassification success rate, and our attacks outperform ADV$^2$ in creating adversarial samples with interpretation maps that are similar to the benign interpretation maps. Furthermore, this level of effectiveness is maintained with very little noise. 

\item We explore the transferability of the attacks across different interpreters. Our results show that perturbation generated using one interpreter can be transferred to another and successfully achieve the adversarial objective in most cases. 

\item We propose two countermeasures based on DNNs and interpreters. We present multiple-interpreter-based detector to identify adversarial samples. We show that the detector improves the adversarial detection process against the proposed attacks. We also present interpretation-based robust training as a defense technique to train a DNN classifier against the worst-case interpretation discrepancy, and indicate that the technique improves the robustness of both a DNN model and its interpreter.

\end{itemize}


\BfPara{Organization} The rest of the paper is organized as follows:
\S \ref{related_work} highlights the relevant literature; \S \ref{fundamentals} presents the fundamental concepts; \S \ref{our_attack} and \S \ref{implementation} provide description of the proposed attacks and the implementation details when used against four interpreters; \S \ref{attack_evaluation} shows the experiments and evaluation of attacks; \S \ref{discussion} presents specific cases in which the existing and our attacks cannot perform well; \S \ref{countermeasure} discusses possible countermeasures; and \S \ref{conclusion} offers our conclusion.

\section{Related Work} \label{related_work}
In this section, we present the related works in three categories: adversarial attacks, transferability, and interpretability.

\BfPara{Adversarial Attacks} Machine learning (ML) models have often become targets for malicious attacks, as they are being increasingly utilized in security-related domains.  In general, adversarial attacks are divided into two primary groups in terms of malicious data manipulations: 1) poisoning the training dataset to degrade the targeted models (\ie poisoning attack \cite{biggio2012poisoning}) and 2) altering the sample input to cause the target model to misbehave (\ie evasion attack \cite{dalvi2004adversarial}). Attacking DNN models has proven to be more complicated because of their complex network architecture as compared with traditional ML models~\cite{lecun2015deep, juraev2022depth}.
There are several pieces of research that focus on attacking DNNs based on new evasion attacks \cite{madry2017towards, szegedy2013intriguing}.
This work particularly investigates attacks against IDLSes that utilize interpretability as defense mechanisms. 

\BfPara{Transferability} Transferability is an interesting property of adversarial attacks \cite{szegedy2013intriguing}: \eg adversarial inputs generated for a specific DNN can be effective against other DNNs. This property is used in black-box scenarios \cite{abdukhamidov2022black}: \eg generating adversarial inputs based on known white-box DNN and applying them against the target black-box DNN \cite{liu2016delving, papernot2016transferability}. In a research study \cite{huang2019enhancing}, a technique that perturbs the hidden layers of a model to improve the transferability of adversarial inputs, is proposed. Another work \cite{xie2019improving} applies differentiable transformations to input samples for enhanced transferability.
In this work, we investigate the transferability of adversarial inputs across different interpreters.

\BfPara{Interpretability} Interpretable models have been utilized to provide the interpretability for DNN models. Such interpretability can be derived using various methods: back-propagation \cite{simonyan2013deep, springenberg2014striving}, intermediate representations \cite{ selvaraju2017grad, zhou2016learning}, input perturbation \cite{fong2017interpretable}, and meta models \cite{dabkowski2017real}.

Interpretability can contribute to system security in the decision-making process involving an expert who can inspect/verify the attribution maps contributing to certain outputs. The interpretation of adversarial inputs is expected to differ remarkably from the interpretation of their benign inputs.
There have been many works on interpretability to debug DNNs \cite{nguyen2015deep}, identifying the adversarial inputs \cite{liu2018adversarial,tao2018attacks}, \etc 

Another recent study \cite{zhang2020interpretable} shows the feasibility of manipulating IDLSes. It offers a novel attacking method to fool DNN models and their corresponding interpretation models simultaneously, demonstrating that the improved interpretability guarantees only limited security and protection.
This work introduces two optimized versions of the ADV$^2$ attack \cite{zhang2020interpretable} against IDLSes to fool the DNN models and mislead the coupled interpretation models. 

\section{Fundamental Concepts} \label{fundamentals}
This section introduces and defines common concepts and terms used throughout the study.
We note that even though this work focuses on IDLSes designed for classification tasks, such as image classification, the attacks can be generalized to other modeling tasks.

Let a classifier $f(x)=y \in Y$ (\ie DNN classifier $f$) that assigns a sample input ($x$) to a category ($y$) from a collection of predetermined categories ($Y$).
Let an interpreter $g(x; f)= m$ (\ie interpretation model $g$) that produces an interpretation map ($m$) that highlights the feature importance in the sample input ($x$) based on the prediction of the DNN model ($f$). The value of the $i\text{-th}$ component in $m$ (\ie $m[i]$) represents the importance of the $i\text{-th}$ component in the sample $x$ (\ie $x[i]$)). 

From this perspective, we consider that model interpretation can be obtained by two basic methods: \cib{1} \textbf{Post-hoc interpretation}: This can be accomplished by adjusting the complexity of DNNs or by employing post-training techniques. This strategy necessitates the development of a new model to provide interpretations for the present model \cite{dabkowski2017real, fong2017interpretable, sabour2017dynamic, simonyan2014deep, zhang2018interpretable}. \cib{2} \textbf{Intrinsic interpretation}: This type of interpretability can be produced by developing self-interpretable DNNs, which intrinsically embed the interpretability feature into their structures \cite{sabour2017dynamic, zhang2018interpretable}.


In this study, we focus on the first method of interpretability, in which an interpreter ($g$) obtains an attribution map $m$ about how a DNN classifier $f$ identifies the sample input $x$.

In a typical adversarial setting, DNN models are vulnerable to adversarial examples, which are maliciously crafted samples aimed to fool the model \cite{moosavi2017universal, szegedy2013intriguing}. 
More specifically, an adversarial input ($\hat{x}$) can be generated by manipulating the benign input ($x$) using one of the well-known attacks (PGD \cite{madry2017towards}, BIM \cite{kurakin2016adversarial}, C\&W \cite{carlini2017towards}, StAdv \cite{xiao2018spatially}, \etc) to make the model misclssify $\hat{x}$ to  an output $f(\hat{x}) = y_{t} \neq f(x)$. 
The manipulations, \eg also known as adversarial perturbations, are usually bounded by a norm ball $\mathcal{B}_{\varepsilon}(x) = \{\left \| \hat{x} - x \right \|_\infty \leq \varepsilon\}$, where $\varepsilon$ is the threshold, to ensure its success and evasiveness. 

In this work, we adopt the following three well-known attack frameworks in our paper: 

\BfPara{PGD} A first-order adversarial attack applies a sequence of project gradient descent on the loss function: 
\begin{equation}\label{eq:pgd}
\begin{split}
    \hat{x}^{(i+1)} = \prod _{\mathcal{B}_{\varepsilon}(x)}\left(\hat{x}^{(i)} - \alpha . ~sign(\nabla_{\hat{x}}\ell_{prd}(f(\hat{x}^{(i)}), y_{t}))\right)
\end{split}
\end{equation}

Here, $\prod$ is a projection operator, $\mathcal{B}_{\varepsilon}$ is a norm ball restrained by a pre-fixed $\varepsilon$, $\alpha$ is a learning rate, $x$ is the benign sample, $\hat{x}^{(i)}$ is the $\hat{x}$ at the iteration $i$, and $\ell_{prd}$ is a loss function that calculates the difference between $f(\hat{x})$ and $y_{t}$.

\BfPara{C\&W} C\&W attack framework considers the process of generating adversarial inputs as a optimization problem of $L_p$-norm regarding the distance metric $\delta$ with respect to the given input $x$, which can be described as:
\begin{equation}\label{eq:c_w_1}
\begin{split}
    ||\mathbf{\delta}||_{p} = \left(\sum_{i=1}^n |\delta_i|^p\right)^{1/p}; \;\; \delta_i = \hat{x}_i - x_i 
\end{split}
\end{equation}
\begin{equation}\label{eq:c_w_2}
\begin{split}
    \text{minimize} \; ||\mathbf{\delta}||_{p} + c \cdot \ell(x + \delta) \;\;\; s.t. \;\;\; x+\delta \in [0, 1]^n, 
\end{split}
\end{equation}

\noindent where $\delta$ is the perturbation added to the input $x$, $\ell$ is a loss function that is to solve optimization problem using gradient descent, and $c$ is a constant number. We utilize $L_2$-norm based C\&W in this work. 

\BfPara{StAdv} The attack is based on the spatial transformation of pixels to improve the perceptual quality of adversarial perturbations. The per-pixel flow is expressed as the flow vector: $r_i := (\Delta u^{(i)}, \Delta v^{(i)})$, specifically, $\Delta u^{(i)} = u_{x}^{(i)} - u_{\hat{x}}^{(i)}$ and $\Delta v^{(i)} = v_{x}^{(i)} - v_{\hat{x}}^{(i)}$, where $(u_{x}^{(i)}, v_{x}^{(i)})$ and $(u_{\hat{x}}^{(i)}, v_{\hat{x}}^{(i)})$ represent the 2D coordinates of the $i$-th pixel of the benign and the adversarial inputs: $x$ and $\hat{x}$. Obtaining the pixel values of the adversarial input is described as:
\begin{equation}\label{eq:stadv_pixel}
\begin{aligned}
    \hat{x}^{(i)} = \sum_{q \in N(u_{x}^{(i)}, v_{x}^{(i)})} x^{(q)} (1 - |u_{x}^{(i)} - u_{x}^{(q)}|) (1 - |v_{x}^{(i)} - v_{x}^{(q)}|),
\end{aligned}
\end{equation}
where $N(u_{x}^{(i)}, v_{x}^{(i)})$ represents the indices of the four pixel neighbors (top/bottom-left/right) of $(u_{x}^{(i)}, v_{x}^{(i)})$. Following two loss functions are used to formulate the search of adversarial deformations as an optimization problem and they are described as follows:
\begin{equation*}\label{eq:stadv_adv_loss}
\begin{split}
    \ell_{prd}(x, r) := \max \left\{ \max_{i \neq t} (f_i(x + r)) - f_t(x + r), \kappa \right\},  
\end{split}
\end{equation*}
\begin{equation*}\label{eq:stadv_flow_loss}
\scalebox{.87}{
$\begin{aligned}
    \ell_{flow}(r) := \sum_{p}^{\text{all\_pixels}} \sum_{q \in N(p)} \sqrt{|| \Delta u^{(p)} - \Delta u^{(q)}||_{2}^2 + || \Delta v^{(p)} - \Delta v^{(q)}||_{2}^2}
\end{aligned}
$}
\end{equation*}

The optimization problem is balanced with a factor of $\lambda$ between the two losses to encourage the adversarial input to be classified as the target label while preserving a high perceptual quality, as follows:
\begin{equation}\label{eq:stadv}
\begin{split}
    \min_{r} \: \left\{\ell_{prd}(x, r) + \lambda \ell_{flow} (r) \right\}
\end{split}
\end{equation}

\section{A\lowercase{dv}E\lowercase{dge} Attack} \label{our_attack}
The section provides the details of implementing the proposed attacks against four different types of interpreters.

\subsection{Attack Definition}
The attack's primary goal is to fool the DNN models $f$ and their coupled interpretation models $g$ in IDLSes. To achieve this goal, an adversarial sample $\hat{x}$ is created by adding noise to the benign version $x$ in order to meet the following conditions:

\begin{enumerate}[leftmargin=2em]

\item A DNN model $f$ misclassifies the adversarial sample $\hat{x}$ to $y_{t}$: $f(\hat{x})=y_{t}$ such that $f(\hat{x}) = y_{t} \neq f(x)$.

\item The adversarial sample $\hat{x}$ causes the interpretation model $g$ to generate the target interpretation map $m_{t}$: $g(\hat{x};f) \approx ~ g(x;f)$.

\item The adversarial and benign samples are indistinguishable.

\end{enumerate}

The proposed approach aims to discover a minimal perturbation, so the adversarial examples satisfy these conditions.
The optimization framework to characterize the attack is:
\begin{equation} \label{eq:nonlinear}
\begin{split}
    \min _{\hat{x}} : \Delta(\hat{x}, x) \, \, s.t. \left\{ \begin{array}{lcr}
         f(\hat{x})=y_{t} \\
         g(\hat{x}; f) = m_{t}
         \end{array}\right.
\end{split}
\end{equation} 

Eq. \ref{eq:nonlinear} can be reformulated to be more appropriate for optimization as follows:

\begin{equation} \label{eq:mainFormula}
\scalebox{0.9}{$
\begin{aligned}
    \min _{\hat{x}}: \ell_{prd}(f(\hat{x}), y_{t}) + \lambda .~\ell_{int}(g(\hat{x};f), m_{t})
    ~s.t. \; \Delta(\hat{x}, x) \leq \varepsilon,
\end{aligned}$}
\end{equation}

\noindent where $\ell_{prd}$ and $\ell_{int}$ denote the classification loss as in Eq. \ref{eq:pgd} and the interpretation loss, respectively. $\ell_{int}$ quantifies the dissimilarity between the adversarial and benign attribution maps, \ie $g(\hat{x}; f)$ and $m_{t} = g(x; f)$, respectively. The hyper-parameter $\lambda$ is used to balance the two components (\ie $\ell_{prd}$ and $\ell_{int}$). 
We build Eq. \ref{eq:mainFormula} based on the PGD framework, and it can be applied to the other attack frameworks with few modifications (\ie considering Eq. \ref{eq:c_w_2} and Eq. \ref{eq:stadv}) as follows:  

\begin{equation} \label{eq:cwMainFormula}
\begin{split}
\scalebox{0.96}{$
    \begin{aligned}
     \min _{\hat{x}} :\; ||\mathbf{\delta}||_{p} + c ~\ell_{prd}(f(\hat{x}), y_{t}) + \lambda .~\ell_{int}(g(\hat{x};f), m_{t})
    \end{aligned}$}
\end{split}
\end{equation} 

\begin{equation} \label{eq:stadvMainFormula}
\begin{split}
\scalebox{0.96}{$
    \begin{aligned}
    \min _{\hat{x}} : \ell_{prd}(f(\hat{x}), y_{t}) + \lambda .~\ell_{int}(g(\hat{x};f), m_{t}) + \lambda \ell_{flow} (r)
    \end{aligned}$}
\end{split}
\end{equation} 

The terms are defined as: 
$$\ell_{prd}(f(\hat{x}), y_{t}) = -~\log(f_{y_{t}}(\hat{x}))\text{,}$$ 
$$\Delta(\hat{x}, x) = \Vert~ \hat{x} - x ~\Vert _{\infty}\text{, and}$$ $$\ell_{int}(g(\hat{x};f), m_{t}) = \Vert g(\hat{x};f) - m_{t} \Vert^{2} _{2}.$$ 

In PGD, the perturbation in Eq. \ref{eq:pgd} is controlled as:
\begin{equation} \label{eq:finalFormula}
\begin{split}
    \hat{x}^{(i+1)} = \prod _{\mathcal{B}_{\varepsilon}(x)}\left(\hat{x}^{(i)} - N_{w}~ \alpha .~ sign(\nabla_{\hat{x}}\ell_{adv}(\hat{x}^{(i)}))\right)
\end{split}
\end{equation}

For C\&W and StAdv attack frameworks, we control the perturbation in Eq. \ref{eq:c_w_1} and Eq. \ref{eq:stadv} as follows:
\begin{equation} \label{eq:cwFinalFormula}
\begin{split}
    \delta = N_{w}~(\hat{x} - x),
\end{split}
\end{equation}
\begin{equation}\label{eq:stadvFinalFormula}
\begin{split}
    r := N_{w}~(\Delta u, \Delta v),
\end{split}
\end{equation}
\noindent where, $N_{w}$ is the perturbation method that determines both the amount and position of noise injected to the sample considering the edge weights $w$ of the benign sample. The entire loss equation (\ie Eq. \ref{eq:mainFormula}) is represented as $\ell_{adv}$. 

\BfPara{\ours}
In Eq. \ref{eq:finalFormula}, we use the $N_{w}$ term to optimize the position and amount of applied perturbation. In \ours{}, we limit the added perturbation to the edges of the sample that overlap with the interpretation map provided by the interpreter. 
In other words, we learn the critical portions of a given input sample based on the overall loss (\ie DNN loss and interpretation model loss) and then inject noise on the edges of those areas.
We define the edges based on the edge weights acquired by applying the Sobel filter \cite{sobelsobel}) to the sample image. 

A pixel-wise edge weights matrix is defined as $\mathcal{E} : e \rightarrow \mathbb{R}^{h \times w}$ for an image of height $h$ and width $w$ using a typical edge extraction method (\eg Sobel filter in our experiments). 
These weights are applied to the sign of the gradient update (PGD framework) as follows: 
  $\mathcal{E}(x) \otimes ~\alpha .~sign(\nabla_{\hat{x}}\ell_{adv}(\hat{x}^{(i)}))$, where the operator $\otimes$ represents the Hadamard product. By doing so, the attack amplifies the noise on the edges while reducing or constricting its presence otherwise in the image.

Considering a straightforward application of the edge weight matrix, \ie $N_{w} = \mathcal{E} (x)$, we can rewrite Eq. \ref{eq:finalFormula}, Eq. \ref{eq:cwFinalFormula}, and Eq. \ref{eq:stadvFinalFormula} (respectively) as:
\begin{equation} \label{eq:finalFormulaAdvEdge}
\begin{split}
    \hat{x}^{(i+1)} = \prod _{\mathcal{B}_{\varepsilon}(x)}\left(\hat{x}^{(i)} - \mathcal{E}(x) ~ \alpha .~  sign(\nabla_{\hat{x}}\ell_{adv}(\hat{x}^{(i)}))\right)
\end{split}
\end{equation}
\begin{equation} \label{eq:cwFinalFormulaAdvEdge}
\begin{split}
    \delta = \mathcal{E}(x) ~(\hat{x} - x)
\end{split}
\end{equation}
\begin{equation}\label{eq:stadvFinalFormulaAdvEdge}
\begin{split}
    r := \mathcal{E}(x) ~(\Delta u, \Delta v)
\end{split}
\end{equation}

\BfPara{\oursplus} This technique, like \ours{}, leverages the edge weights to optimize the perturbation. Instead of adjusting the noise based on the edge weights in the attack, we apply the perturbations only on defined edges that pass a certain threshold.  
This is performed by a binarization operation (\eg $bin(\mathcal{E}_\delta(x) \rightarrow e_i := \{0 ~\text{if}~ e_i \leq \delta;~1 ~\text{otherwise} ~\forall ~ i\})$) to obtain a binary edge matrix ${\mathcal{E}_\delta} : e \rightarrow [0, 1]^{h \times w}$, where $h$ and $w$ are the height and width of an input respectively. The hyperparameter $\delta$ sets the threshold for binarizing edge weights, and can be adjusted based on the attack objective.
The rest of the implementation is similar to \ours{}, \eg  
the PGD attack can be formulated as: 
  ${\mathcal{E}_\delta}(x) \otimes ~\alpha .~sign(\nabla_{\hat{x}}\ell_{adv}(\hat{x}^{(i)}))$.
This enables perturbations to be applied only on the edges.
Given the constraint on perturbation locality and amount, the threshold $\delta$ is adjusted to $0.1$ to improve the effectiveness.

The implementation of attacks in Equations (\ref{eq:finalFormula}, \ref{eq:cwFinalFormula}, \ref{eq:stadvFinalFormula}, \ref{eq:finalFormulaAdvEdge}, \ref{eq:cwFinalFormulaAdvEdge}, and \ref{eq:stadvFinalFormulaAdvEdge}) against the representatives of four categories of interpreters is discussed in the following subsections: 1) \textit{back-propagation-guided interpretation}, 2) \textit{representation-guided interpretation}, 3) \textit{model-guided interpretation}, and 4) \textit{perturbation-guided interpretation}.

\subsection{Back-Propagation-Guided Interpretation}
This category of interpretation models computes the gradient of a DNN model's prediction with respect to the provided sample input. This highlights the significance of each feature in the input sample. Larger values suggest higher importance of the features to the model prediction. In this paper, we discuss gradient saliency (Grad) \cite{simonyan2014deep} as an example of back-propagation-guided interpretation. The Grad interpreter generates an attribution map $m$ given the model $f$ prediction to a given input $x$ with a given class $y$. This attribution map is estimated as follows: 
\begin{equation*} \label{eq:gradFormula}
\begin{split}
    m = \left| \frac{\partial f_{y} (x)}{\partial x} \right|
\end{split}
\end{equation*}

Looking for the optimal adversarial sample $\hat{x}$ for IDLSes based on Grad via a sequence of gradient descent updates (as in applying Eq. \ref{eq:finalFormula}) is ineffective, since DNN models with ReLU activation functions cause the Hessian matrix computation result to be all-zero. 
The challenge can be resolved by computing the smoothed value of the gradient of ReLU ($r(z)$) using a function $h(z)$ with the following formula:

\begin{equation*} \label{eq:smoothRelu}
\begin{split}
    h(z) \triangleq \left\{ \begin{array}{lcr}
         (z+\sqrt{z^{2}+\tau})' = 1+\frac{z}{\sqrt{z^{2}+\tau}} & \mbox{for}
         & z < 0 \\
         (\sqrt{z^{2} + \tau})' = \frac{z}{\sqrt{z^{2} + \tau}} & \mbox{for} & z \geq 0
                \end{array}\right. 
\end{split}
\end{equation*}

Here, $\tau$ is a constant parameter, and $h(z)$ approximates the values of $r(z)$ and its gradients are non-zero. Another alternative is to use the sigmoid function ($\sigma(z) = \frac{1}{1+e^{-z}}$).
We note that this method can be applied to other back-propagation-guided interpreters ( \eg LRP \cite{bach2015pixel}, SmoothGrad \cite{smilkov2017smoothgrad}) as they are based on gradient-centric formulations \cite{ancona2017towards}.    

\subsection{Representation-Guided Interpretation}

This type of interpreter extracts feature maps from intermediate layers of DNNs to provide attribution maps.
The representative of representation-guided interpretation in this study is \textbf{C}lass \textbf{A}ctivation \textbf{M}ap (CAM) \cite{zhou2016learning}.
CAM performs global average pooling on the convolutional feature maps, and utilizes the outputs as features for a fully-connected layer for the model prediction. The significance of the sample input areas can be determined by projecting the output layer weights back onto the convolutional feature maps.

Specifically, let $a_{i}(k, l)$ be the activation of channel $i$ in the last convolutional layer at $(k, l)$ spatial location, and $\sum_{k, l} a_{i}(k, l)$ as the result of global average pooling. The activation map $m_{c}$ for an input $x$ and class $c$ is given by:
\begin{equation*} \label{eq:camFormula}
\begin{split}
    m_{c}(x, y) = \sum_{i} w_{i, c} a_{i} (k, l),
\end{split}
\end{equation*}
\noindent where $c$ is $c$-th output of the linear layer and $w_{i, c}$ is the weight corresponding to class $c$ for $i$-th channel. 
We construct $m$ by extracting and concatenating interpretation maps from $f$ up to the final convolutional layer and a fully connected layer, similar to the work of \cite{zhang2020interpretable}. We exploit the interpretation model by utilizing gradient descent updates to find the optimal $\hat{x}$ as in Eq. \ref{eq:finalFormula}. The attack can also be used against various interpreters of this category (\eg Grad-CAM \cite{selvaraju2017grad}). 

\subsection{Model-Guided Interpretation}
By masking significant areas in any input, a model-guided interpretation model is trained to predict the attribution map in a single forward pass directly. We consider \textbf{R}eal-\textbf{T}ime Image \textbf{S}aliency (RTS) \cite{dabkowski2017real} for this category.
RTS generates the attribution map $m$ for any given class $c$ and input $x$ by resolving the following problem:
\begin{equation} \label{eq:rtsFormula}
\begin{aligned}
    \min_{m} :  & \quad\lambda_{1}r_{tv}(m) + \lambda_{2} r_{av}(m) \\
    & - \log(f_{c}(\Phi(x; m))) \\
                & +\lambda_{3}f_{c}(\Phi(x; 1-m))^{\lambda_{4}} \;\; s.t. \; 0 \leq m \leq 1
\end{aligned}
\end{equation}

Here, the regularizers $\{\lambda_{i}\}_{i=1}^4$  balance the involved factors, the $r_{av}(m)$ represents the average value of $m$. The $r_{tv}(m)$ expresses the variation of $m$ that enforces mask smoothness, and it is defined simply as:
$r_{tv}(m) = \sum_{i, j}(m_{i,j} - m_{i,j + 1})^2 + \sum_{i, j}(m_{i,j} - m_{i+1, j})^2$. Furthermore, $\Phi$ is the operator that uses $m$ as a mask to blend $x$ with random color noise and Gaussian blur, and it is defined as:
$\Phi(x, m) = x \otimes m + r \otimes (1 - m)$, 
where $\otimes$ denotes Hadamard product and $r$ is an alternative image. For applying $\Phi$, $r$ can be chosen to be a highly blurred version of in the input $x$. In general, Eq. \ref{eq:rtsFormula} finds the important parts of $x$ in terms of which $f$ predicts $f(x)$ with high confidence.

During the inference, computing Eq. \ref{eq:rtsFormula} is expensive. Therefore, RTS trains a DNN to predict the attribution map for any input so that RTS does not have to access the DNN $f$ after training. 
This can be achieved by composing a ResNet \cite{he2016deep}, a pre-trained model as an encoder to extract feature maps of the given inputs and a U-Net \cite{ronneberger2015u} as the masking model which is trained to optimize the framework in Eq. \ref{eq:rtsFormula}. For this work, we consider the composition of the encoder and masking model as the interpreter $g$.

Attacking RTS by employing Eq. \ref{eq:finalFormula}, Eq. \ref{eq:cwFinalFormula} or Eq. \ref{eq:stadvFinalFormula} directly, has been found to be inefficient for finding the optimal adversarial samples \cite{zhang2020interpretable}. The main reason for this is that the interpretation model uses both the masking model and the encoder ($enc(\cdot)$).
To address this issue, we utilize an additional loss term $\ell_{enc}(enc(\hat{x}), enc(y_{t}))$ to Eq. \ref{eq:mainFormula}, Eq. \ref{eq:cwMainFormula} and Eq. \ref{eq:stadvMainFormula} to compute the difference between the encoder's outcome with the adversarial sample input $\hat{x}$ and the target class $y_{t}$. Then, we perform a series of gradient descent updates to choose the best the optimal adversarial sample input $\hat{x}$.

\subsection{Perturbation-Guided Interpretation}
This type of interpretation model investigates the interpretation maps by introducing a minimal amount of noise to the input and evaluating the changes in the model's prediction. We consider MASK \cite{fong2017interpretable} to be the class representative.
 
MASK identifies the attribution map for the given input $x$ by affecting the maximally informative input regions. Specifically, the model $f(x)$ generates a vector of scores for distinct hypotheses about the input $x$ (\eg as a softmax probability layer in a DNN), then MASK explores the smallest mask $m$ to make the model performance drop significantly: \ie $f_{y}(\Phi(x; m)) \leqslant f_{y}(x)$.
We note that the mask $m$ in this scenario is binary, where $m[i] = 0 ~\text{or}~ 1$ to indicate whether the $i\text{-th}$ feature is replaced with Gaussian noise.
The optimal mask can be obtained by solving the following problem: 
\begin{equation} \label{eq:maskFormula1}
\begin{split}
     \min_{m} : f_{y}(\Phi(x; m)) + \lambda . ~\Vert 1 - m \Vert _{1}  \;\; s.t. \; 0 \leq m \leq 1,
\end{split}
\end{equation}
\noindent where $\lambda$ encourages most of the mask to be sparse. By solving the Eq. \ref{eq:maskFormula1}, we find the most informative and necessary regions of the input $x$ with reference to the model prediction $f(x)$. 

We cannot directly advance the Eq. \ref{eq:mainFormula}, Eq. \ref{eq:cwMainFormula} and Eq. \ref{eq:stadvMainFormula} with the Eq. \ref{eq:finalFormula}, Eq. \ref{eq:cwFinalFormula} and Eq. \ref{eq:stadvFinalFormula}, respectively since the interpretation model $g$ is built as an optimization method.  

\sloppy A bi-level optimization framework \cite{zhang2020interpretable} can be used to solve this problem. The loss function is rewritten as follows: 
$\ell_{adv}(x, m)$ $\triangleq$ $\ell_{prd}(f(x), y_{t}) + \lambda .~ \ell_{int}(m, m_{t})$ by including  benign attribution map $m$ as a new variable.
Let $\ell_{map}(m;x)$ denote the object function of Eq. \ref{eq:maskFormula1} and $m_{\ast}(x) = \argmin_{m} : \ell_{map}(m; x)$ be the attribution map generated by MASK for the input $x$. Then, we have the following framework:
\begin{equation} \label{eq:maskFormula2}
\scalebox{0.94}{$
\begin{aligned}
     & \min_{x} : \ell_{adv}(x, m_{\ast}(x)) \, 
      s.t. \; m_{\ast}(x) = \argmin_{m} \, \ell_{map}(m;x)
\end{aligned}$}
\end{equation}

For every update on the input $x$, solving the bi-level optimization in Eq. \ref{eq:maskFormula2} is expensive. As a solution for the issue, an approximate iterative procedure was proposed by \cite{zhang2020interpretable} to optimize $x$ and the mask $m$ by alternating between $\ell_{adv}$ and $\ell_{map}$.  
Briefly, given $x^{(i-1)}$ at the $i$-th iteration, the attribution map $m^{(i)}$ is computed by updating $m^{(i-1)}$ on $\ell_{map}(m^{(i-1)}; x^{(i-1)})$, then $m^{(i)}$ is fixed and $x^{(i)}$ is attained by reducing $\ell_{adv}$ after one gradient descent step with respect to $m^{(i)}$. To update $x^{(i)}$, the objective function is formulated as follows:
\begin{equation} \label{eq:maskFormula3}
\begin{split}
     \ell_{adv}\left(x^{(i-1)}, m^{(i)} - \alpha .~ \nabla_{m} .~ \ell_{map}(m^{(i)}; x^{(i-1)})\right),
\end{split}
\end{equation}
\noindent where, $\alpha$ is the learning rate. Implementation details are provided in the following section.
\begin{figure}[t]
    \centering
    \captionsetup{justification=justified}
    \includegraphics[width=0.49\textwidth]{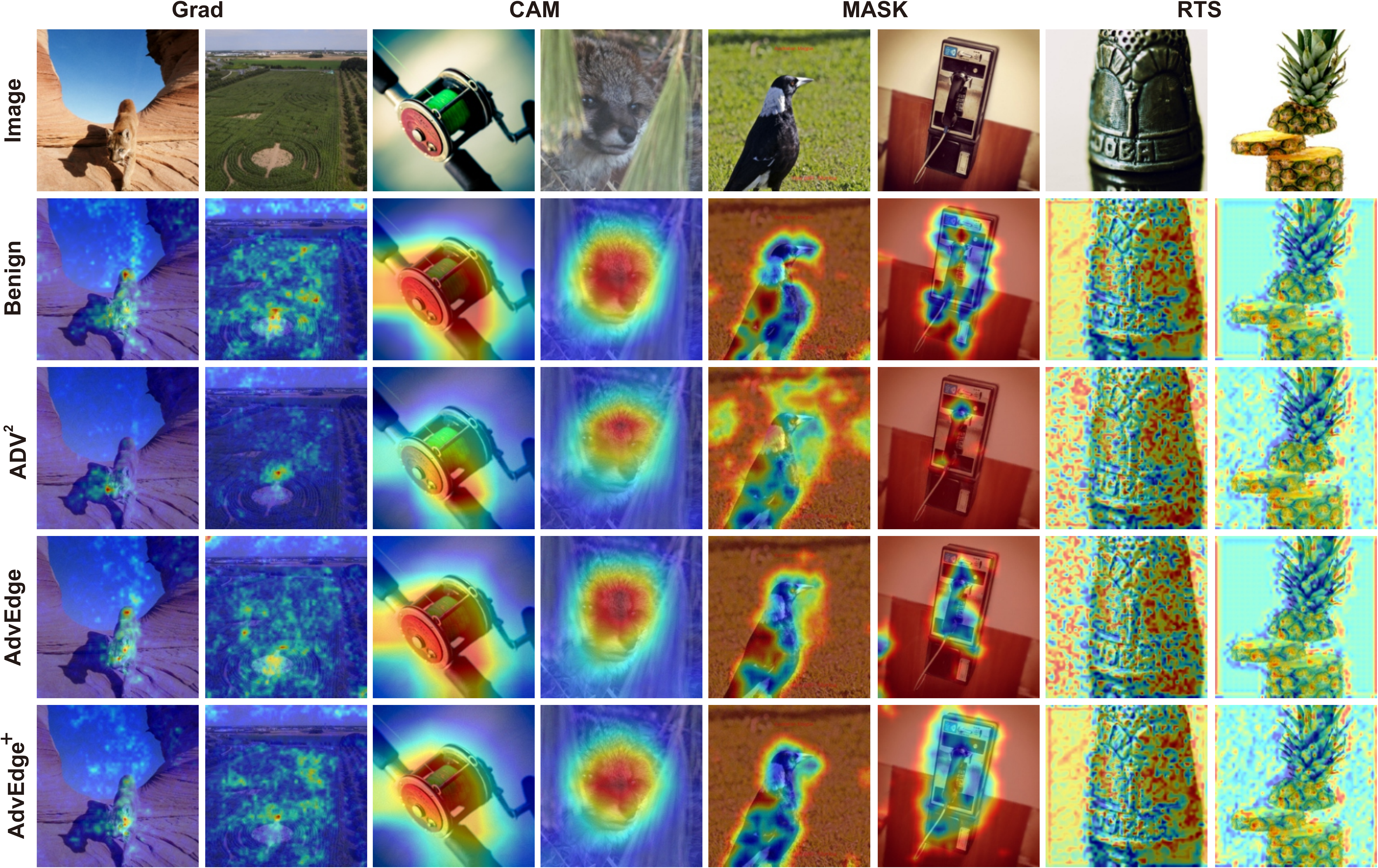}
    \caption{Examples of benign and adversarial attribution maps generated by Grad, CAM, MASK, and RTS interpreters on the ResNet model. Adversarial attribution maps are based on ADV$^2$, \ours, and \oursplus attacks.}
    \label{fig:sample_set}
\end{figure}

\section{Experimental Setting} \label{implementation}
This section describes the implementation details of  \ours{} and \oursplus{}, and the methods used to maximize their effectiveness on target interpreters (Grad, CAM, MASK, and RTS).
Both \ours and \oursplus are built based on the PGD, C\&W and StAdv attack frameworks. We use the values $\alpha = {1/255}$, and $\varepsilon = 0.031$ from the prior work \cite{zhang2020interpretable}. ${\ell_{\infty}}$ is used to calculate the perturbation proportion. To improve the attack's efficiency, we use a method that adds noise to the edges of the images with a fixed number of iterations ($\#\text{iterations} =300$). 


\BfPara{Optimization Steps} zero gradient of the prediction loss hinders finding the desired result with correct interpretation (Grad). To address this issue, a label smoothing strategy using cross-entropy is used. The approach samples the prediction loss using a uniform distribution $\mathbb{U}(1 - \rho, 1)$ and the value of $\rho$ is substantially decreased during the attacking process. Considering 
$y_{c} = \frac{1-y_{t}}{|Y|-1}$, we derive $\ell_{prd}(f(x), y_{t})=-\sum_{c \in Y} y_{c} \log f_{c}(x)$.

In solving the Eq. \ref{eq:maskFormula3}, multiple steps of gradient descent to update $m$ and to compute the $m_{\ast}(x)$ for faster convergence are applied. Additionally, the average gradient is utilized to update $m$ and to support the stable optimization. More precisely, at the $i$-th iteration with multistep gradient descent, $\{m_{j}^{(i)}\}$ is the sequence of maps. To calculate the gradient in order to update $m$, the interpretation loss $\sum_{j} \Vert m_{j}^{(i)} - m_{t} \Vert _{2} ^{2}$ is used. To improve the convergence, the learning rate is also changed dynamically. 
For stable learning, $x$ is updated in two steps: \cib{1} updating $x$ based on the prediction loss $\ell_{prd}$; \cib{2} updating $x$ based on the interpretation loss $\ell_{int}$. 

In the second step, $x$'s confidence score is checked whether it is still above a specific threshold after the perturbation by searching the largest step size with applying the binary search. Updating the estimation of the attribution map in terms of $\ell_{map}$ results in the significant deviation from the original map generated by MASK as the number of steps increases in the update process. To keep the effectiveness of the attack, the generated map is replaced with the map $g(x;f)$ which is the output of the MASK in terms of the latest adversarial input at a specific iteration point (\eg 50 iterations). Simultaneously, Adam step parameter is reset for the correct state.        

\BfPara{Dataset} For our experiment, we use ImageNetV2 Top-Images \cite{recht2019imagenet} dataset. ImageNetV2 is a new test set collected based on the ImageNet benchmark and was mainly published for inference accuracy evaluation. The dataset contains 10,000 images based on 1,000 different classes that are similar to the classes in the original ImageNet dataset. All images are cropped to 224 $\times$ 224 pixels, and the pixels are normalized between $[0, 1]$. As the test set, we use 1,000 images that are selected randomly and classified correctly by the classifier $f$. 

\BfPara{Prediction Models} The study employs two state-of-the-art DNN models \textbf{ResNet-50} \cite{he2016deep} and \textbf{DenseNet-169} \cite{huang2017densely}, which demonstrate 22.85\% and 22.08\% top-1 error rates on the ImageNet dataset, respectively. Those DNN models differ in capacity (50 and 169 layers, respectively) and network architecture (residual blocks and dense blocks, respectively). Using those various DNN models assist in computing the efficacy of the proposed attacks. 


\BfPara{Interpretation Models} We explored the attacks against the Grad \cite{simonyan2014deep}, CAM \cite{zhou2016learning}, RTS \cite{dabkowski2017real}, and MASK \cite{fong2017interpretable} interpreters as representative of back-propagation-guided, representation-guided, model-guided, and perturbation-guided interpretation models, respectively. In our experiment, we employed the interpreters' original open-source implementation.

\BfPara{Attack Framework} We implement our attack on the basis of PGD, C\&W, and StAdv attack frameworks. We note that attack frameworks other than the used in this work (\eg BIM \cite{kurakin2016adversarial}) can also be implemented for the attack. 
We also assume that our proposed attacks are based on {non-targeted attacks}, in which DNN models are forced to misclassify the adversarial input $\hat{x}$ to a certain and randomly-selected class. We make comparisons between the proposed approaches (\ours and \oursplus) and ADV$^2$ attack, which is assumed a novel attack to produce adversarial samples for target DNNs and their linked interpreters. We have the same hyperparameters (\textit{learning rate}, \textit{number of iterations}, \textit{step size}, \etc) and experimental settings as in ADV$^2$ \cite{zhang2020interpretable}.

\section{Attack Evaluation} \label{attack_evaluation}

For the evaluation, we answer the following questions:
\cib{1} Are the \ours and \oursplus attacks effective against DNN models?
\cib{2} Are the \ours and \oursplus attacks efficient at deceiving interpreters?
\cib{3} Does transferability exist in the attacks against the interpreters using the proposed attacks?
\cib{4} Do the \ours and \oursplus attacks make it easier to attack IDLSes?
To answer these questions, we perform several experiments evaluating \ours and \oursplus, and comparing the findings to $\text{ADV}^{2}$ in~\cite{zhang2020interpretable}. We utilize the original implementation of $\text{ADV}^{2}$ presented by the authors.

\BfPara{Evaluation Metrics} We used several evaluation metrics to assess the attack success against the used DNN classifiers and interpretation models. To begin, we use the following metrics to evaluate the attacks based on misleading the DNN models: 

\begin{itemize}[leftmargin=1em]
    \item \textbf{Attack success rate:} We calculate the proportion of successfully-misclassified test inputs to the total number of test samples, which is calculated as: $\frac{\#\text{successful trials}}{\#\text{total trials}}$.
    \item \textbf{Misclassification confidence:} We check confidence in predicting the targeted category, which is the probability given to the class $y_{t}$ by a DNN model.
\end{itemize}

Furthermore, we assess the attacks based on their capability in deceiving the interpretation model.
This is accomplished by assessing the interpretation maps of adversarial samples. This task is challenging since there is a lack of standard metrics to evaluate the interpretation maps provided by the interpreters. Therefore, we use the following metrics:

\begin{itemize}[leftmargin=1em]
    \item \textbf{Qualitative comparison}: We use this measurement to check whether the results of our approach are perceptually indistinguishable. We compare the interpretations maps of benign and adversarial inputs qualitatively.
    \item \textbf{$\mathcal{L}_{p}$ Measure}: We observe the difference between benign and adversarial attribution maps using the $\mathcal{L}_{1}$ distance. All values of attribution maps are standardized to $[0, 1]$ to obtain the results.
    \item \textbf{IoU Test} (\textbf{I}ntersection-\textbf{o}ver-\textbf{U}nion): This is another quantitative metric for determining the similarity of attribution maps. This metric is commonly used to compare the prediction outcome with the ground truth:
    IoU($m$) = $| O(m) \bigcap O(m_{\circ})| ~ / ~| O(m) \bigcup  O(m_{\circ}) |$, where $m$ is adversarial attribution map and $m_{\circ}$ is the benign attribution map. This metric is also known as Jaccard index.    
\end{itemize}


\begin{table*}[t]
\centering
\caption{Attack success rate of ADV$^2$, \ours, and \oursplus against different classifiers (ResNet-50 and DenseNet-169) and interpreters (Grad, CAM, RTS, and MASK) testing on 1,000 images and using PGD, C\&W, and stAdv frameworks.}
\label{tab:ASR}
\resizebox{\linewidth}{!}{%
\begin{tabular}{|c|c|c|c|c|c|c|c|c|c|c|c|c|c|} 
\hline
\multirow{2}{*}{}               & \multirow{2}{*}{} & \multicolumn{3}{c|}{\textbf{Grad}}                                                       & \multicolumn{3}{c|}{\textbf{CAM}}                                                        & \multicolumn{3}{c|}{\textbf{MASK}}                                                       & \multicolumn{3}{c|}{\textbf{RTS}}                                                         \\ 
\hhline{|~~------------|}
                                &                   & {\cellcolor[rgb]{0.718,0.718,0.718}}\textbf{ADV$^2$} & \textbf{AdvEdge} & \textbf{AdvEdge$^+$} & {\cellcolor[rgb]{0.718,0.718,0.718}}\textbf{ADV$^2$} & \textbf{AdvEdge} & \textbf{AdvEdge$^+$} & {\cellcolor[rgb]{0.718,0.718,0.718}}\textbf{ADV$^2$} & \textbf{AdvEdge} & \textbf{AdvEdge$^+$} & {\cellcolor[rgb]{0.718,0.718,0.718}}\textbf{ADV$^2$} & \textbf{AdvEdge} & \textbf{AdvEdge$^+$}  \\ 
\hline
\multirow{2}{*}{\textbf{PGD}}   & {\cellcolor[rgb]{0.922,0.902,0.902}}\textbf{Resnet}   & {\cellcolor[rgb]{0.718,0.718,0.718}}1.00          & {\cellcolor[rgb]{0.922,0.902,0.902}}1.00             & {\cellcolor[rgb]{0.922,0.902,0.902}}1.00              & {\cellcolor[rgb]{0.718,0.718,0.718}}1.00          & {\cellcolor[rgb]{0.922,0.902,0.902}}1.00             & {\cellcolor[rgb]{0.922,0.902,0.902}}1.00              & {\cellcolor[rgb]{0.718,0.718,0.718}}1.00          & {\cellcolor[rgb]{0.922,0.902,0.902}}1.00             & {\cellcolor[rgb]{0.922,0.902,0.902}}1.00              & {\cellcolor[rgb]{0.718,0.718,0.718}}1.00          & {\cellcolor[rgb]{0.922,0.902,0.902}}1.00             & {\cellcolor[rgb]{0.922,0.902,0.902}}1.00               \\ 
\hhline{|~-------------|}
                                & \textbf{Densenet} & {\cellcolor[rgb]{0.718,0.718,0.718}}1.00          & 1.00             & 1.00              & {\cellcolor[rgb]{0.718,0.718,0.718}}1.00          & 1.00             & 1.00              & {\cellcolor[rgb]{0.718,0.718,0.718}}1.00          & 1.00             & 1.00              & {\cellcolor[rgb]{0.718,0.718,0.718}}1.00          & 1.00             & 1.00               \\ 
\hline
\multirow{2}{*}{\textbf{C\&W}}  & {\cellcolor[rgb]{0.922,0.902,0.902}}\textbf{Resnet}   & {\cellcolor[rgb]{0.718,0.718,0.718}}1.00          & {\cellcolor[rgb]{0.922,0.902,0.902}}1.00             & {\cellcolor[rgb]{0.922,0.902,0.902}}1.00              & {\cellcolor[rgb]{0.718,0.718,0.718}}1.00          & {\cellcolor[rgb]{0.922,0.902,0.902}}1.00             & {\cellcolor[rgb]{0.922,0.902,0.902}}1.00              & {\cellcolor[rgb]{0.718,0.718,0.718}}1.00          & {\cellcolor[rgb]{0.922,0.902,0.902}}1.00             & {\cellcolor[rgb]{0.922,0.902,0.902}}1.00              & {\cellcolor[rgb]{0.718,0.718,0.718}}0.99          & {\cellcolor[rgb]{0.922,0.902,0.902}}0.99             & {\cellcolor[rgb]{0.922,0.902,0.902}}0.99               \\ 
\hhline{|~-------------|}
                                & \textbf{Densenet} & {\cellcolor[rgb]{0.718,0.718,0.718}}1.00          & 0.99             & 0.99              & {\cellcolor[rgb]{0.718,0.718,0.718}}0.94          & 0.94             & 0.94              & {\cellcolor[rgb]{0.718,0.718,0.718}}1.00          & 1.00             & 1.00              & {\cellcolor[rgb]{0.718,0.718,0.718}}0.99          & 0.99             & 0.99               \\ 
\hline
\multirow{2}{*}{\textbf{StAdv}} & {\cellcolor[rgb]{0.922,0.902,0.902}}\textbf{Resnet}   & {\cellcolor[rgb]{0.718,0.718,0.718}}0.99          & {\cellcolor[rgb]{0.922,0.902,0.902}}0.99             & {\cellcolor[rgb]{0.922,0.902,0.902}}0.99              & {\cellcolor[rgb]{0.718,0.718,0.718}}1.00          & {\cellcolor[rgb]{0.922,0.902,0.902}}1.00             & {\cellcolor[rgb]{0.922,0.902,0.902}}1.00              & {\cellcolor[rgb]{0.718,0.718,0.718}}0.99          & {\cellcolor[rgb]{0.922,0.902,0.902}}0.96             & {\cellcolor[rgb]{0.922,0.902,0.902}}0.95              & {\cellcolor[rgb]{0.718,0.718,0.718}}0.99          & {\cellcolor[rgb]{0.922,0.902,0.902}}0.99             & {\cellcolor[rgb]{0.922,0.902,0.902}}0.99               \\ 
\hhline{|~-------------|}
                                & \textbf{Densenet} & {\cellcolor[rgb]{0.718,0.718,0.718}}0.97          & 0.97             & 0.96              & {\cellcolor[rgb]{0.718,0.718,0.718}}0.94          & 0.94             & 0.94              & {\cellcolor[rgb]{0.718,0.718,0.718}}0.99          & 0.98             & 0.97              & {\cellcolor[rgb]{0.718,0.718,0.718}}0.99          & 0.99             & 0.99               \\
\hline
\end{tabular}}
\end{table*}


\begin{table*}[t]
\centering
\caption{Misclassification confidence of ADV$^2$, \ours, and \oursplus against different classifiers (ResNet-50 and DenseNet-169) and interpreters (Grad, CAM, RTS, and MASK) testing on 1,000 images and using PGD, C\&W, and stAdv.}
\label{tab:MC}
\arrayrulecolor{black}
\resizebox{\linewidth}{!}{%
\begin{tabular}{|c|c|c|c|c|c|c|c|c|c|c|c|c|c|} 
\hline
\multirow{2}{*}{}               & \multirow{2}{*}{} & \multicolumn{3}{c|}{\textbf{Grad}}                                                       & \multicolumn{3}{c|}{\textbf{CAM}}                                                        & \multicolumn{3}{c|}{\textbf{MASK}}                                                       & \multicolumn{3}{c|}{\textbf{RTS}}                                                         \\ 
\hhline{|~~------------|}
                                &                   & {\cellcolor[rgb]{0.718,0.718,0.718}}\textbf{ADV$^2$} & \textbf{AdvEdge} & \textbf{AdvEdge$^+$} & {\cellcolor[rgb]{0.718,0.718,0.718}}\textbf{ADV$^2$} & \textbf{AdvEdge} & \textbf{AdvEdge$^+$} & {\cellcolor[rgb]{0.718,0.718,0.718}}\textbf{ADV$^2$} & \textbf{AdvEdge} & \textbf{AdvEdge$^+$} & {\cellcolor[rgb]{0.718,0.718,0.718}}\textbf{ADV$^2$} & \textbf{AdvEdge} & \textbf{AdvEdge$^+$}  \\ 
\hline
\multirow{2}{*}{\textbf{PGD}}   & {\cellcolor[rgb]{0.922,0.902,0.902}}\textbf{Resnet}   & {\cellcolor[rgb]{0.718,0.718,0.718}}0.927         & {\cellcolor[rgb]{0.922,0.902,0.902}}\textbf{0.936}   & {\cellcolor[rgb]{0.922,0.902,0.902}}0.923             & {\cellcolor[rgb]{0.718,0.718,0.718}}0.553         & {\cellcolor[rgb]{0.922,0.902,0.902}}\textbf{0.554}   & {\cellcolor[rgb]{0.922,0.902,0.902}}\textbf{0.560}    & {\cellcolor[rgb]{0.718,0.718,0.718}}0.535         & {\cellcolor[rgb]{0.922,0.902,0.902}}\textbf{0.597}   & {\cellcolor[rgb]{0.922,0.902,0.902}}0.534             & {\cellcolor[rgb]{0.718,0.718,0.718}}0.715         & {\cellcolor[rgb]{0.922,0.902,0.902}}0.705            & {\cellcolor[rgb]{0.922,0.902,0.902}}\textbf{0.717}     \\ 
\hhline{|~-------------|}
                                & \textbf{Densenet} & {\cellcolor[rgb]{0.718,0.718,0.718}}0.903         & \textbf{0.921}   & \textbf{0.910}    & {\cellcolor[rgb]{0.718,0.718,0.718}}0.501         & 0.491            & \textbf{0.512}    & {\cellcolor[rgb]{0.718,0.718,0.718}}0.585         & \textbf{0.631}   & \textbf{0.628}    & {\cellcolor[rgb]{0.718,0.718,0.718}}0.566         & \textbf{0.584}   & \textbf{0.574}     \\ 
\hline
\multirow{2}{*}{\textbf{C\&W}}  & {\cellcolor[rgb]{0.922,0.902,0.902}}\textbf{Resnet}   & {\cellcolor[rgb]{0.718,0.718,0.718}}0.925         & {\cellcolor[rgb]{0.922,0.902,0.902}}\textbf{0.930}   & {\cellcolor[rgb]{0.922,0.902,0.902}}\textbf{0.931}    & {\cellcolor[rgb]{0.718,0.718,0.718}}0.833         & {\cellcolor[rgb]{0.922,0.902,0.902}}\textbf{0.835}   & {\cellcolor[rgb]{0.922,0.902,0.902}}\textbf{0.840}    & {\cellcolor[rgb]{0.718,0.718,0.718}}0.401         & {\cellcolor[rgb]{0.922,0.902,0.902}}\textbf{0.410}   & {\cellcolor[rgb]{0.922,0.902,0.902}}0.397             & {\cellcolor[rgb]{0.718,0.718,0.718}}0.614         & {\cellcolor[rgb]{0.922,0.902,0.902}}\textbf{0.665}   & {\cellcolor[rgb]{0.922,0.902,0.902}}\textbf{0.644}     \\ 
\hhline{|~-------------|}
                                & \textbf{Densenet} & {\cellcolor[rgb]{0.718,0.718,0.718}}0.930         & 0.929            & \textbf{0.930}    & {\cellcolor[rgb]{0.718,0.718,0.718}}0.789         & \textbf{0.790}   & \textbf{0.792}    & {\cellcolor[rgb]{0.718,0.718,0.718}}0.439         & \textbf{0.482}   & \textbf{0.450}    & {\cellcolor[rgb]{0.718,0.718,0.718}}0.593         & \textbf{0.596}   & \textbf{0.593}     \\ 
\hline
\multirow{2}{*}{\textbf{StAdv}} & {\cellcolor[rgb]{0.922,0.902,0.902}}\textbf{Resnet}   & {\cellcolor[rgb]{0.718,0.718,0.718}}0.885         & {\cellcolor[rgb]{0.922,0.902,0.902}}0.877            & {\cellcolor[rgb]{0.922,0.902,0.902}}0.870             & {\cellcolor[rgb]{0.718,0.718,0.718}}0.947         & {\cellcolor[rgb]{0.922,0.902,0.902}}\textbf{0.950}   & {\cellcolor[rgb]{0.922,0.902,0.902}}\textbf{0.949}    & {\cellcolor[rgb]{0.718,0.718,0.718}}0.401         & {\cellcolor[rgb]{0.922,0.902,0.902}}\textbf{0.405}   & {\cellcolor[rgb]{0.922,0.902,0.902}}0.397             & {\cellcolor[rgb]{0.718,0.718,0.718}}0.839         & {\cellcolor[rgb]{0.922,0.902,0.902}}\textbf{0.843}   & {\cellcolor[rgb]{0.922,0.902,0.902}}\textbf{0.844}     \\ 
\hhline{|~-------------|}
                                & \textbf{Densenet} & {\cellcolor[rgb]{0.718,0.718,0.718}}0.866         & 0.853            & 0.849             & {\cellcolor[rgb]{0.718,0.718,0.718}}0.870         & \textbf{0.883}   & \textbf{0.873}    & {\cellcolor[rgb]{0.718,0.718,0.718}}0.439         & \textbf{0.482}   & \textbf{0.446}    & {\cellcolor[rgb]{0.718,0.718,0.718}}0.769         & \textbf{0.771}   & \textbf{0.773}     \\
\hline
\end{tabular}}
\end{table*}

The following metric is used to quantify the amount of noise to produce the adversarial inputs:
\begin{itemize}[leftmargin=1em]
    \item \textbf{Structural Similarity (SSIM)}: The mean structural similarity index \cite{wang2004image} between the benign and adversarial sample inputs is used to calculate added the noise. SSIM is a technique to measure the image quality based on its distortion-free reference image:
    \begin{equation*} \label{eq:SSIM}
        \begin{split}
            \text{SSIM}(x, \hat{x}) =  \frac{(2 \mu_{x} \mu_{\hat{x}} + c_{1})(2 \sigma_{x\hat{x}} + c_{2})}{(\mu^{2} _{x} + \mu^{2} _{\hat{x}} + c_{1})(\sigma^{2} _{x} + \sigma^{2} _{\hat{x}} + c_{2})},
        \end{split}
    \end{equation*}
    where $\mu_{x}$ and $\mu_{\hat{x}}$ are the averages of $x$ and $\hat{x}$, $\sigma_{x}$ and $\sigma_{\hat{x}}$ are the variances of $x$ and $\hat{x}$, $\sigma_{x\hat{x}}$ is the covariance of $x$ and $\hat{x}$. $c_{1}$ and $c_{2}$ are the variables to stabilize the division with weak denominator. To calculate the non-similarity rate (also known as the distance or noise rate), we subtract the SSIM value from one ($\text{noise\_rate} = 1 - \text{SSIM}$).
    
\end{itemize}

Finally, we also consider the time for the attacks:
\begin{itemize}[leftmargin=1em]
    \item \textbf{Average Time}: We calculate the time that is taken to generate adversarial inputs by the attacks. 
\end{itemize}


\subsection{Attack Effectiveness against DNNs}
In terms of fooling the target DNN models, we first evaluate \ours and \oursplus along with comparing the results with the previous work (ADV$^2$ \cite{zhang2020interpretable}).
We build existing method upon C\&W and StAdv frameworks for comparison purposes. The results are shown in Tables \ref{tab:ASR} and \ref{tab:MC}, and presented using \textit{attack success rate} and \textit{misclassification confidence}.

Table \ref{tab:ASR} summarizes the results of the attack success rate of ADV$^2$, \ours, and \oursplus against various DNN models and interpreters based on 1,000 images. As it is shown, the attack success rate for C\&W and StAdv is not as high as PGD. The reason for the case is that we applied the same hyperparameter values of PGD attack for the other attacks as fair comparison(\textit{step size}, \textit{number of iterations}, \textit{learning rate}, \etc). In general, \ours and \oursplus are as effective as ADV$^2$. Additionally, Table \ref{tab:MC} presents the results of misclassification confidence on 1,000 images in terms of three methods against various DNN and interpretation models.
Despite the fact that the main goal is to apply a minimal amount of noise to certain parts of the sample images, the performance on three attack frameworks is quite better than ADV$^2$ in terms of Grad (ResNet), CAM (DenseNet) and RTS (DenseNet) in three attack frameworks.
In other circumstances, the confidence levels of the DNN models are comparable. The results show that \ours and \oursplus are effective in deceiving the target DNNs. Restricting perturbations to edges, the proposed methods are as effective as ADV$^2$.

\begin{figure*}
 \begin{minipage}{\textwidth}
  \centering
  \includegraphics[width=0.9\textwidth]{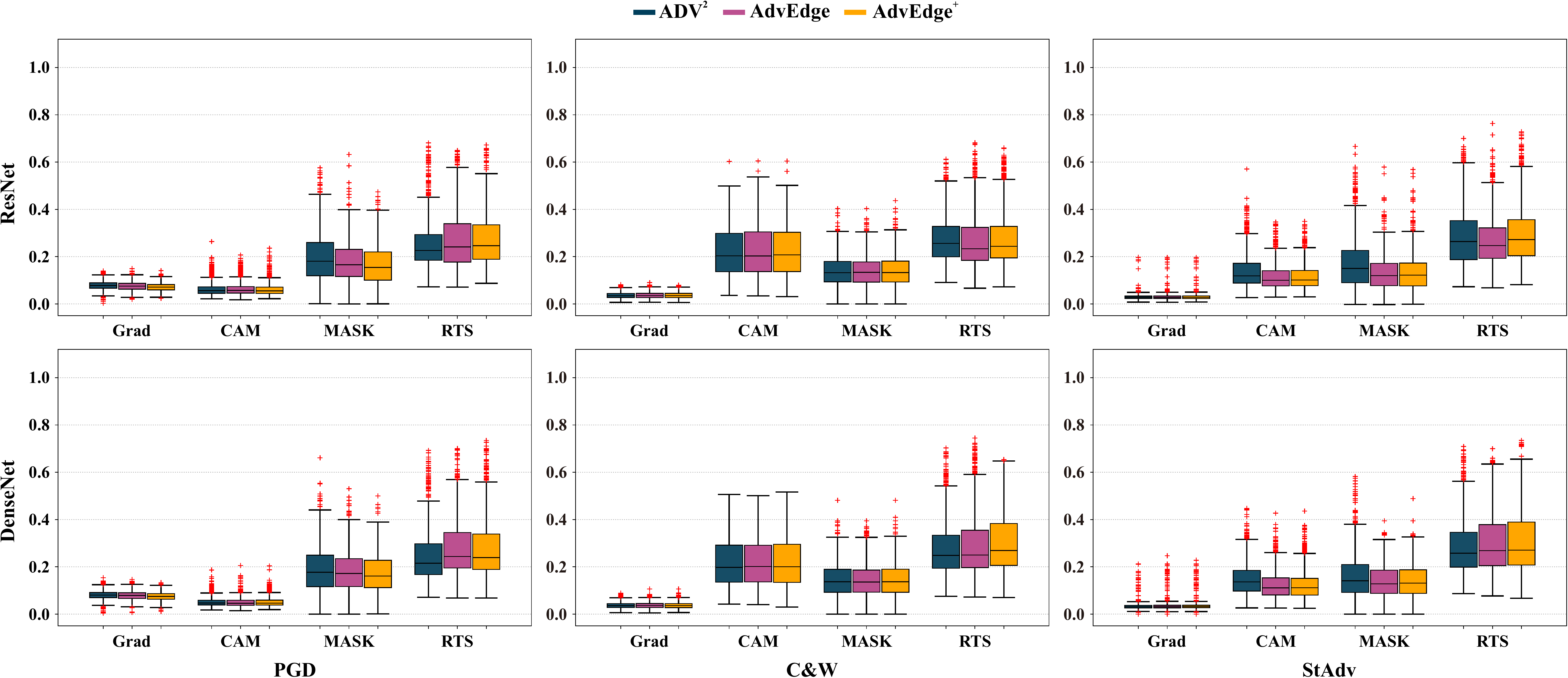}
  \caption{$\mathcal{L}_{1}$ distance of attribution maps generated by different interpreters when applying ADV$^2$, \ours, and \oursplus from those of corresponding benign samples on ResNet-50 and DenseNet-169 using the PGD, C\&W, and stAdv frameworks.}
 \label{fig:lp_all}
 \end{minipage}
\end{figure*}

\begin{figure*}[t]
    \centering
    \captionsetup{justification=justified}
    \includegraphics[width=0.9\textwidth]{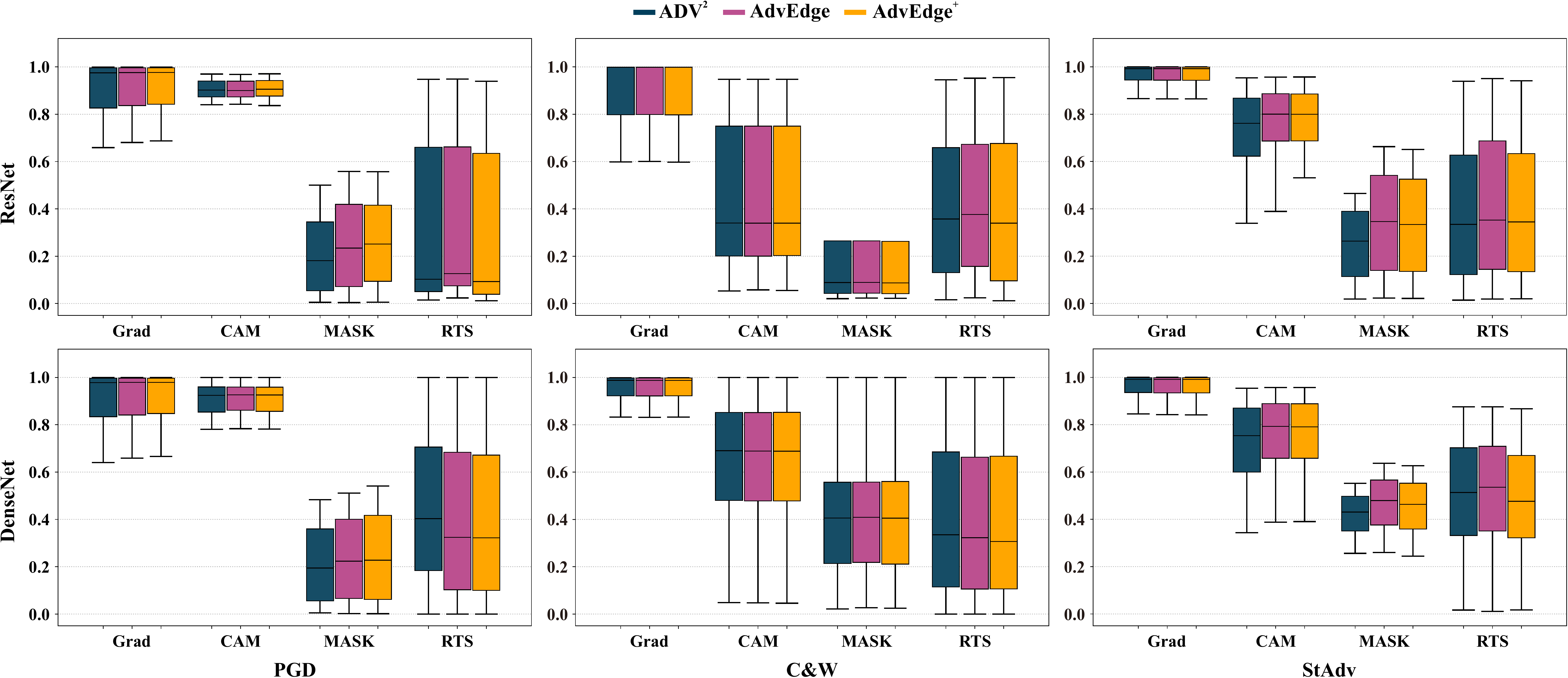}
    \caption{IoU scores of attribution maps generated by ADV$^2$, \ours, and \oursplus using the four interpreters on ResNet-50 and DenseNet-169. Our attacks achieve higher IoU scores in comparison with ADV$^2$. }
    \label{fig:iouscore}
\end{figure*}

\subsection{Attack Effectiveness against Interpreters}
We assess the capability of \ours and \oursplus attacks to create interpretation maps that are comparable to the benign interpretation maps. We begin with a qualitative evaluation to observe the similarity of interpretations produced by adversarial samples (\ie generated by \ours and \oursplus) and the corresponding benign samples. By making qualitative comparison of the interpretations of benign and adversarial samples, \ours and \oursplus generated interpretations that are visually indistinguishable from their corresponding benign sample inputs. When compared to ADV$^2$, both proposed approaches produced interpretation maps that are similar to the benign inputs. Figure \ref{fig:sample_set} shows some examples of observed attribution maps obtained using Grad, CAM, RTS, and MASK. The examples show the adversarial interpretations and the corresponding benign. As shown in the figure, the adversarial and benign attribution maps are highly similar.

In addition to qualitative analysis, we employ $\mathcal{L}_{p}$ to quantify the similarity of generated interpretation maps. Figure \ref{fig:lp_all} summarizes the results of $\mathcal{L}_{1}$ measurement. As displayed in the figure, when compared to ADV$^2$, \ours and \oursplus produce adversarial sample inputs with interpretation maps that are similar to those obtained for the benign samples. The results are consistent among interpretation models with all DNNs. We observe that the efficiency of our attack (against interpretation models) differs with the interpretation models. Furthermore, as reflected in the Figure \ref{fig:lp_all}, there are several cases that the attacks could not achieve high attack success rate in terms of interpreters, which are known as outliers. We will analyse them in \S \ref{discussion}.

The IoU score is another quantitative metric for comparing the similarity of interpretation maps. We binarized the values of interpretation maps to compute the IoU as they are originally real values.
Figure \ref{fig:iouscore} displays the IoU scores of attribution maps generated by ADV$^2$, \ours, and \oursplus adopting PGD, C\&W, and StAdv with regard to four interpreter models on ResNet and DenseNet. As shown in the figure, \ours and \oursplus performed better than ADV$^2$. We observe that \ours and \oursplus achieved significantly better results in terms of the interpreters Grad, CAM, and MASK. \ours and \oursplus are effective in deceiving different interpreters. Based upon the qualitative and quantitative measures, the attribution maps of adversarial and benign samples are almost indistinguishable.


\begin{figure*}[t]
\centering
\includegraphics[width=0.98\textwidth]{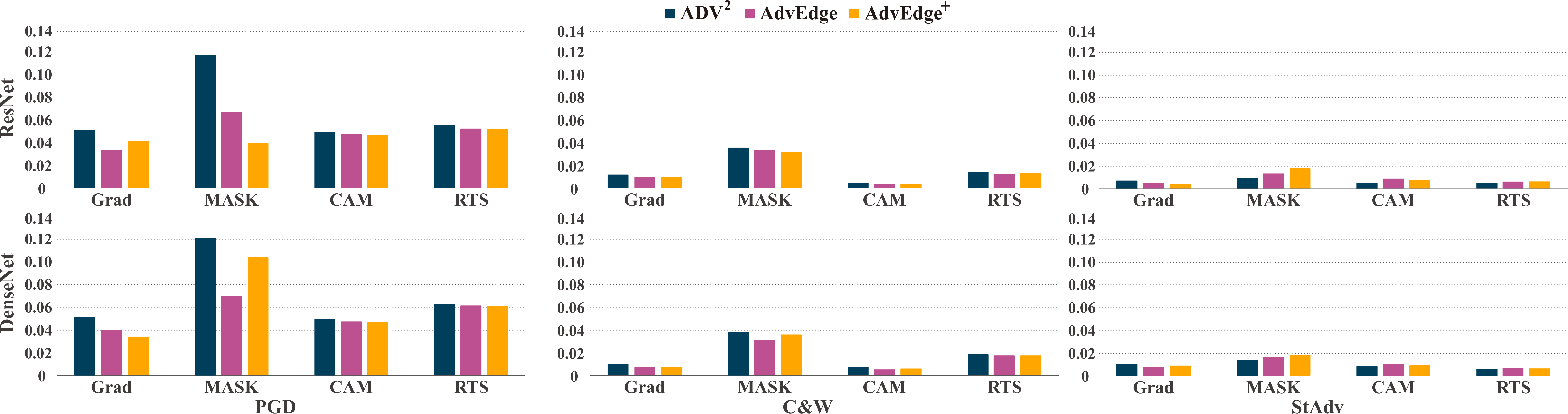}
\caption{Noise rate of adversarial inputs generated by ADV$^2$, \ours, and \oursplus on ResNet-50 and DenseNet-169.}
\label{fig:noise_rate}
\end{figure*}

\begin{table*}[t]
\centering
\caption{Average time (in seconds) to generate a single adversarial input by ADV$^2$, \ours and \oursplus across different interpreters on ResNet-50 and DenseNet-169. Best results are given in bold.}
\label{tab:average_time}
\arrayrulecolor{black}
\resizebox{\linewidth}{!}{%
\begin{tabular}{!{\color{black}\vrule}c!{\color{black}\vrule}c!{\color{black}\vrule}c!{\color{black}\vrule}c!{\color{black}\vrule}c!{\color{black}\vrule}c!{\color{black}\vrule}c!{\color{black}\vrule}c!{\color{black}\vrule}c!{\color{black}\vrule}c!{\color{black}\vrule}c!{\color{black}\vrule}c!{\color{black}\vrule}c!{\color{black}\vrule}c!{\color{black}\vrule}} 
\hline
\multirow{2}{*}{}               & \multirow{2}{*}{} & \multicolumn{3}{c!{\color{black}\vrule}}{\textbf{Grad}}                                  & \multicolumn{3}{c!{\color{black}\vrule}}{\textbf{CAM}}                                   & \multicolumn{3}{c!{\color{black}\vrule}}{\textbf{MASK}}                                  & \multicolumn{3}{c!{\color{black}\vrule}}{\textbf{RTS}}                                    \\ 
\hhline{>{\arrayrulecolor{black}}|~~>{\arrayrulecolor{black}}------------|}
                                &                   & {\cellcolor[rgb]{0.718,0.718,0.718}}\textbf{ADV$^2$} & \textbf{AdvEdge} & \textbf{AdvEdge$^+$} & {\cellcolor[rgb]{0.718,0.718,0.718}}\textbf{ADV$^2$} & \textbf{AdvEdge} & \textbf{AdvEdge$^+$} & {\cellcolor[rgb]{0.718,0.718,0.718}}\textbf{ADV$^2$} & \textbf{AdvEdge} & \textbf{AdvEdge$^+$} & {\cellcolor[rgb]{0.718,0.718,0.718}}\textbf{ADV$^2$} & \textbf{AdvEdge} & \textbf{AdvEdge$^+$}  \\ 
\hline
\multirow{2}{*}{\textbf{PGD}}   & {\cellcolor[rgb]{0.922,0.902,0.902}}\textbf{Resnet}   & {\cellcolor[rgb]{0.718,0.718,0.718}}32.68        & {\cellcolor[rgb]{0.922,0.902,0.902}}32.69           & {\cellcolor[rgb]{0.922,0.902,0.902}}\textbf{30.02}   & {\cellcolor[rgb]{0.718,0.718,0.718}}15.37        & {\cellcolor[rgb]{0.922,0.902,0.902}}16.05           & {\cellcolor[rgb]{0.922,0.902,0.902}}15.83            & {\cellcolor[rgb]{0.718,0.718,0.718}}188.97       & {\cellcolor[rgb]{0.922,0.902,0.902}}\textbf{188.14} & {\cellcolor[rgb]{0.922,0.902,0.902}}194.56           & {\cellcolor[rgb]{0.718,0.718,0.718}}8.75         & {\cellcolor[rgb]{0.922,0.902,0.902}}\textbf{8.45}   & {\cellcolor[rgb]{0.922,0.902,0.902}}\textbf{8.39}     \\ 
\hhline{>{\arrayrulecolor{black}}|~>{\arrayrulecolor{black}}-------------|}
                                & \textbf{Densenet} & {\cellcolor[rgb]{0.718,0.718,0.718}}33.02        & \textbf{33.01}  & \textbf{31.45}   & {\cellcolor[rgb]{0.718,0.718,0.718}}49.42        & \textbf{49.38}  & \textbf{49.37}   & {\cellcolor[rgb]{0.718,0.718,0.718}}340.49       & 343.40          & 340.53           & {\cellcolor[rgb]{0.718,0.718,0.718}}12.43        & \textbf{12.37}  & 16.73    \\ 
\hline
\multirow{2}{*}{\textbf{C\&W}}  & {\cellcolor[rgb]{0.922,0.902,0.902}}\textbf{Resnet}   & {\cellcolor[rgb]{0.718,0.718,0.718}}75.95        & {\cellcolor[rgb]{0.922,0.902,0.902}}\textbf{75.68}  & {\cellcolor[rgb]{0.922,0.902,0.902}}\textbf{75.64}   & {\cellcolor[rgb]{0.718,0.718,0.718}}4.68         & {\cellcolor[rgb]{0.922,0.902,0.902}}\textbf{4.68}            & {\cellcolor[rgb]{0.922,0.902,0.902}}4.72             & {\cellcolor[rgb]{0.718,0.718,0.718}}174.18       & {\cellcolor[rgb]{0.922,0.902,0.902}}185.32          & {\cellcolor[rgb]{0.922,0.902,0.902}}\textbf{139.37}  & {\cellcolor[rgb]{0.718,0.718,0.718}}8.78         & {\cellcolor[rgb]{0.922,0.902,0.902}}8.96            & {\cellcolor[rgb]{0.922,0.902,0.902}}8.95              \\ 
\hhline{>{\arrayrulecolor{black}}|~>{\arrayrulecolor{black}}-------------|}
                                & \textbf{Densenet} & {\cellcolor[rgb]{0.718,0.718,0.718}}173.70       & \textbf{173.09} & \textbf{173.26}  & {\cellcolor[rgb]{0.718,0.718,0.718}}4.46         & \textbf{4.46}            & 4.54             & {\cellcolor[rgb]{0.718,0.718,0.718}}201.81       & \textbf{201.78} & 202.91           & {\cellcolor[rgb]{0.718,0.718,0.718}}12.29        & \textbf{12.23}  & 12.56             \\ 
\hline
\multirow{2}{*}{\textbf{StAdv}} & {\cellcolor[rgb]{0.922,0.902,0.902}}\textbf{Resnet}   & {\cellcolor[rgb]{0.718,0.718,0.718}}80.16        & {\cellcolor[rgb]{0.922,0.902,0.902}}\textbf{79.88}  & {\cellcolor[rgb]{0.922,0.902,0.902}}\textbf{79.96}   & {\cellcolor[rgb]{0.718,0.718,0.718}}13.13        & {\cellcolor[rgb]{0.922,0.902,0.902}}\textbf{13.12}  & {\cellcolor[rgb]{0.922,0.902,0.902}}13.17            & {\cellcolor[rgb]{0.718,0.718,0.718}}987.88       & {\cellcolor[rgb]{0.922,0.902,0.902}}\textbf{731.46} & {\cellcolor[rgb]{0.922,0.902,0.902}}\textbf{732.35}  & {\cellcolor[rgb]{0.718,0.718,0.718}}7.48         & {\cellcolor[rgb]{0.922,0.902,0.902}}7.68            & {\cellcolor[rgb]{0.922,0.902,0.902}}7.82              \\ 
\hhline{>{\arrayrulecolor{black}}|~>{\arrayrulecolor{black}}-------------|}
                                & \textbf{Densenet} & {\cellcolor[rgb]{0.718,0.718,0.718}}179.03       & 179.17          & 179.52           & {\cellcolor[rgb]{0.718,0.718,0.718}}8.14         & 8.29            & 8.24             & {\cellcolor[rgb]{0.718,0.718,0.718}}293.12       & 309.89          & 311.12           & {\cellcolor[rgb]{0.718,0.718,0.718}}10.23        & 10.72           & 10.40             \\
\hline
\end{tabular}
}
\arrayrulecolor{black}
\end{table*}

\subsection{Adversarial Perturbation Rate}
The SSIM metric helps assess the pixel-level perturbation applied by \ours and \oursplus to generate the adversarial samples. We measure the noise amount from images' areas that are not identical to the original images using SSIM.
Figure \ref{fig:noise_rate} presents the perturbation amount generated by different attacks (\ie ADV$^2$, \ours, and \oursplus) to succeed under different frameworks. 
As shown, \ours and \oursplus required significantly lower noise than  ADV$^2$ in terms of PGD and C\&W. 
The results become more obvious when using the MASK interpreter (PGD attack). 
\oursplus generates the least amount of noise to fool the target DNN model and the MASK interpreter. In terms of StAdv, limiting the perturbation for only edges of specific regions caused the attack to use more noise to have successful attack as compared with ADV$^2$. This can be more noticeable, when MASK interpreter is used. Generally, \ours and \oursplus achieve high success rate with less perturbation.



\begin{figure*}[t]
 \begin{minipage}{\textwidth}
  \centering
  \includegraphics[width=0.95\textwidth]{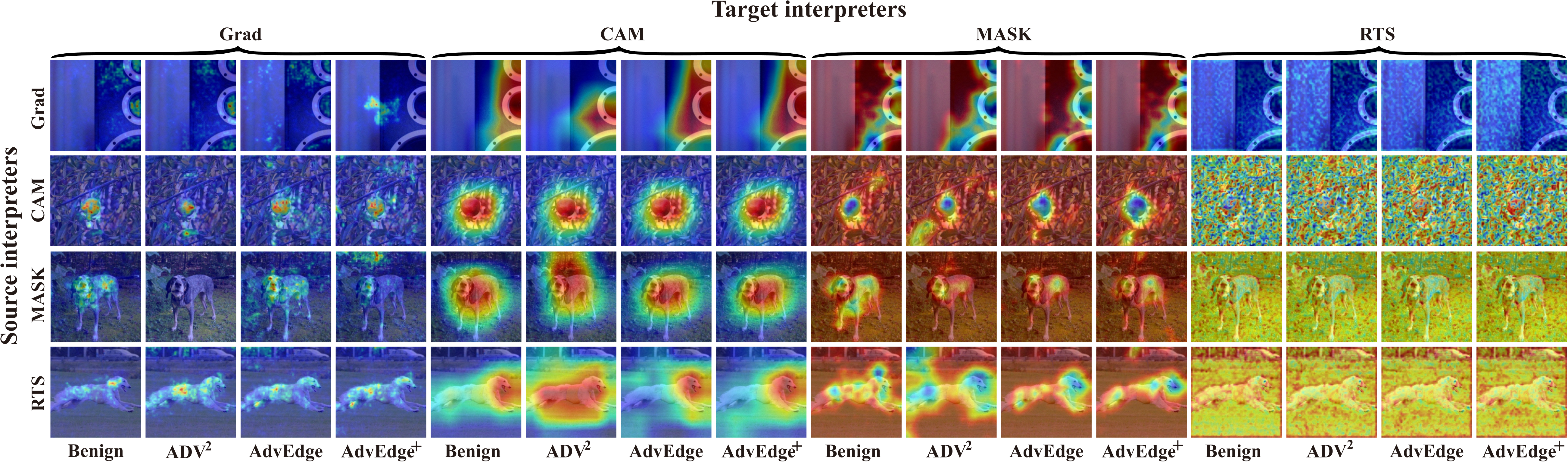}
  \caption{Visualization of transferability for attribution maps of adversarial inputs across different interpreters on ResNet-50 adopting PGD attack framework. Heatmap areas and shapes of the results generated by \ours and \oursplus share high similarity with the benign cases. Our attacks provide significantly better transferability as compared with ADV$^2$.}\vspace{1em}
\label{fig:transferability_example}
 \end{minipage} 
 \begin{minipage}{\textwidth}
  \centering
  \includegraphics[width=0.95\textwidth]{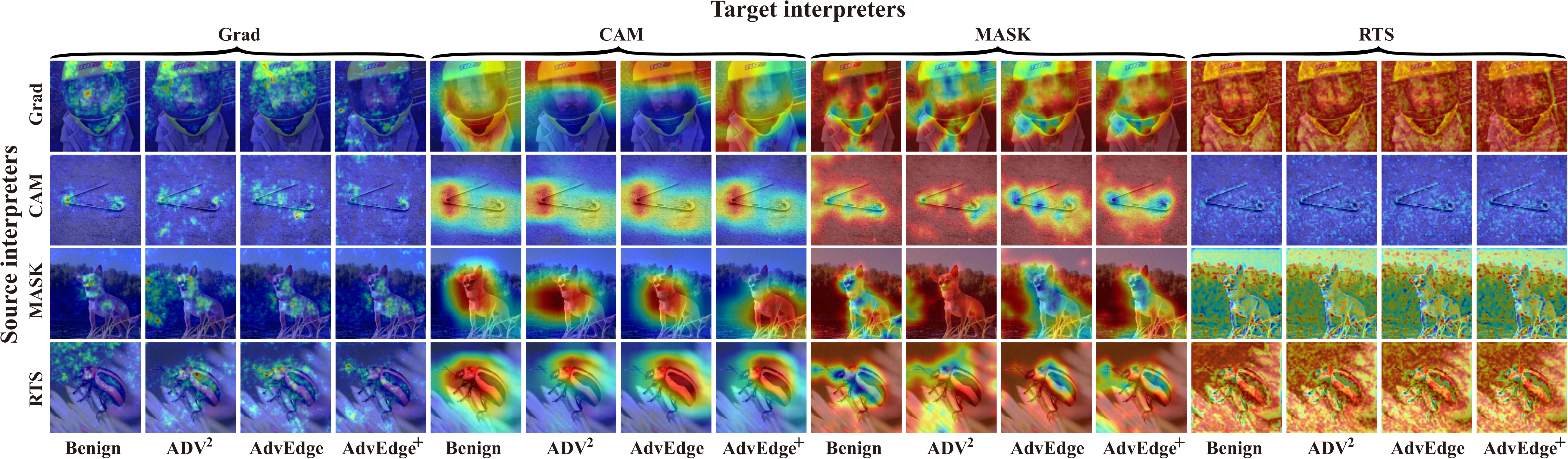}
  \caption{Limitations in transferability among attribution maps of adversarial inputs across different interpreters on ResNet. Our attack results still provide meaningful interpretation in attack transferability.}
    \label{fig:transferability_limitation}
 \end{minipage}
\end{figure*}

\subsection{Average Time}

To compare the attacks in terms of time, we calculated the time to generate adversarial inputs. As shown in Table \ref{tab:average_time}, CAM and RTS interpreters took less amount of time for the generation process in comparison of Grad and MASK in general. Among the four interpreters, MASK is the interpreter that takes considerable time to achieve successful attack rate. In terms of attack frameworks adopted (PGD, StAdv, and C\&W), C\&W based on CAM and MASK is faster while PGD based on Grad generates adversarial in shorter time period. In accordance with the attack types, there is not any significant time difference among ADV$^2$, \ours and \oursplus. In several cases, our attacks took less time to search in adversarial space (Grad). This can clearly be observed in the time taken by \ours and  \oursplus using StAdv on MASK interpreter and ResNet classifier. To conclude, \ours and \oursplus provide comparable time to craft adversarial inputs.

\subsection{Attack Transferability}
One interesting property of adversarial inputs is their transferability. Specifically, one adversarial input effective against a DNN can be effective against other DNNs \cite{liu2016delving, moosavi2017universal, papernot2016transferability}. In this part, we evaluate the transferability in our attacks against interpreters. 
Specifically, for a given interpreter $g$, a set of adversarial inputs generated against $g$ is selected randomly to compute their attribution maps using the other interpreters $g'$. Figure \ref{fig:transferability_example} illustrates the attribution maps of adversarial inputs on $g$ and $g'$ on the basis of PGD attack. We provide the example for PGD attack, as it adds a considerable amount of noise to the inputs for high attack success, which is highly likely for attribution maps to deviate from their original attribution maps (see Figure \ref{fig:noise_rate}). Furthermore, we compare the attribution maps of adversarial inputs against their corresponding original attribution maps. As shown in Figure \ref{fig:transferability_example}, interpretation transferability is valid for the attacks, and it is more obvious in \ours and \oursplus than in ADV$^2$.

\begin{table*}
\centering
\caption{$\mathcal{L}_{1}$ distance of transferability among attribution maps of adversarial (ADV$^2$, \ours, and \oursplus) across interpreters on ResNet-50 and DenseNet-169 (row/column as source/target interpreters). Best results are given in bold.}
\label{tab:l1_transferability}
\resizebox{\linewidth}{!}{%
\begin{tabular}{|c|c|c|c|c|c|c|c|c|c|c|c|c|c|c|} 
\hline
\multirow{2}{*}{}               & \multirow{2}{*}{}                  & \multirow{2}{*}{} & \multicolumn{3}{c|}{\textbf{Grad}}                                                       & \multicolumn{3}{c|}{\textbf{CAM}}                                                        & \multicolumn{3}{c|}{\textbf{MASK}}                                                       & \multicolumn{3}{c|}{\textbf{RTS}}                                                         \\ 
\hhline{|~~~------------|}
                                &                                    &                   & {\cellcolor[rgb]{0.718,0.718,0.718}}\textbf{ADV$^2$} & \textbf{AdvEdge} & \textbf{AdvEdge$^+$} & {\cellcolor[rgb]{0.718,0.718,0.718}}\textbf{ADV$^2$} & \textbf{AdvEdge} & \textbf{AdvEdge$^+$} & {\cellcolor[rgb]{0.718,0.718,0.718}}\textbf{ADV$^2$} & \textbf{AdvEdge} & \textbf{AdvEdge$^+$} & {\cellcolor[rgb]{0.718,0.718,0.718}}\textbf{ADV$^2$} & \textbf{AdvEdge} & \textbf{AdvEdge$^+$}  \\ 
\hline
\multirow{8}{*}{\textbf{PGD}}   & \multirow{4}{*}{\textbf{Resnet}}   & {\cellcolor[rgb]{0.922,0.902,0.902}}\textbf{Grad}     & {\cellcolor[rgb]{0.718,0.718,0.718}}0.081         & {\cellcolor[rgb]{0.922,0.902,0.902}}\textbf{0.077}   & {\cellcolor[rgb]{0.922,0.902,0.902}}\textbf{0.074}    & {\cellcolor[rgb]{0.718,0.718,0.718}}0.246         & {\cellcolor[rgb]{0.922,0.902,0.902}}\textbf{0.241}   & {\cellcolor[rgb]{0.922,0.902,0.902}}\textbf{0.241}    & {\cellcolor[rgb]{0.718,0.718,0.718}}0.578         & {\cellcolor[rgb]{0.922,0.902,0.902}}\textbf{0.563}   & {\cellcolor[rgb]{0.922,0.902,0.902}}\textbf{0.556}    & {\cellcolor[rgb]{0.718,0.718,0.718}}0.092         & {\cellcolor[rgb]{0.922,0.902,0.902}}\textbf{0.077}   & {\cellcolor[rgb]{0.922,0.902,0.902}}\textbf{0.080}     \\ 
\hhline{|~~-------------|}
                                &                                    & \textbf{CAM}      & {\cellcolor[rgb]{0.718,0.718,0.718}}0.093         & \textbf{0.092}   & \textbf{0.091}    & {\cellcolor[rgb]{0.718,0.718,0.718}}0.063         & \textbf{0.063}   & \textbf{0.062}    & {\cellcolor[rgb]{0.718,0.718,0.718}}0.523         & \textbf{0.523}   & \textbf{0.523}    & {\cellcolor[rgb]{0.718,0.718,0.718}}0.097         & \textbf{0.097}   & \textbf{0.096}     \\ 
\hhline{|~~-------------|}
                                &                                    & {\cellcolor[rgb]{0.922,0.902,0.902}}\textbf{MASK}     & {\cellcolor[rgb]{0.718,0.718,0.718}}0.117         & {\cellcolor[rgb]{0.922,0.902,0.902}}\textbf{0.116}   & {\cellcolor[rgb]{0.922,0.902,0.902}}\textbf{0.103}    & {\cellcolor[rgb]{0.718,0.718,0.718}}0.246         & {\cellcolor[rgb]{0.922,0.902,0.902}}\textbf{0.234}   & {\cellcolor[rgb]{0.922,0.902,0.902}}0.262             & {\cellcolor[rgb]{0.718,0.718,0.718}}0.545         & {\cellcolor[rgb]{0.922,0.902,0.902}}\textbf{0.529}   & {\cellcolor[rgb]{0.922,0.902,0.902}}\textbf{0.482}    & {\cellcolor[rgb]{0.718,0.718,0.718}}0.099         & {\cellcolor[rgb]{0.922,0.902,0.902}}\textbf{0.087}   & {\cellcolor[rgb]{0.922,0.902,0.902}}\textbf{0.069}     \\ 
\hhline{|~~-------------|}
                                &                                    & \textbf{RTS}      & {\cellcolor[rgb]{0.718,0.718,0.718}}0.097         & \textbf{0.096}   & \textbf{0.096}    & {\cellcolor[rgb]{0.718,0.718,0.718}}0.281         & \textbf{0.273}   & \textbf{0.273}    & {\cellcolor[rgb]{0.718,0.718,0.718}}0.626         & \textbf{0.613}   & \textbf{0.617}    & {\cellcolor[rgb]{0.718,0.718,0.718}}0.098         & \textbf{0.097}   & \textbf{0.097}     \\ 
\hhline{|~--------------|}
                                & \multirow{4}{*}{\textbf{Densenet}} & {\cellcolor[rgb]{0.922,0.902,0.902}}\textbf{Grad}     & {\cellcolor[rgb]{0.718,0.718,0.718}}0.085         & {\cellcolor[rgb]{0.922,0.902,0.902}}\textbf{0.082}   & {\cellcolor[rgb]{0.922,0.902,0.902}}\textbf{0.079}    & {\cellcolor[rgb]{0.718,0.718,0.718}}0.219         & {\cellcolor[rgb]{0.922,0.902,0.902}}0.222            & {\cellcolor[rgb]{0.922,0.902,0.902}}0.220             & {\cellcolor[rgb]{0.718,0.718,0.718}}0.606         & {\cellcolor[rgb]{0.922,0.902,0.902}}\textbf{0.596}   & {\cellcolor[rgb]{0.922,0.902,0.902}}\textbf{0.575}    & {\cellcolor[rgb]{0.718,0.718,0.718}}0.086         & {\cellcolor[rgb]{0.922,0.902,0.902}}\textbf{0.072}   & {\cellcolor[rgb]{0.922,0.902,0.902}}\textbf{0.073}     \\ 
\hhline{|~~-------------|}
                                &                                    & \textbf{CAM}      & {\cellcolor[rgb]{0.718,0.718,0.718}}0.102         & \textbf{0.101}   & \textbf{0.102}    & {\cellcolor[rgb]{0.718,0.718,0.718}}0.093         & \textbf{0.092}   & \textbf{0.092}    & {\cellcolor[rgb]{0.718,0.718,0.718}}0.573         & \textbf{0.563}   & \textbf{0.564}    & {\cellcolor[rgb]{0.718,0.718,0.718}}0.100         & \textbf{0.099}   & \textbf{0.099}     \\ 
\hhline{|~~-------------|}
                                &                                    & {\cellcolor[rgb]{0.922,0.902,0.902}}\textbf{MASK}     & {\cellcolor[rgb]{0.718,0.718,0.718}}0.117         & {\cellcolor[rgb]{0.922,0.902,0.902}}0.118            & {\cellcolor[rgb]{0.922,0.902,0.902}}\textbf{0.115}    & {\cellcolor[rgb]{0.718,0.718,0.718}}0.250         & {\cellcolor[rgb]{0.922,0.902,0.902}}\textbf{0.246}   & {\cellcolor[rgb]{0.922,0.902,0.902}}0.253             & {\cellcolor[rgb]{0.718,0.718,0.718}}0.546         & {\cellcolor[rgb]{0.922,0.902,0.902}}\textbf{0.523}   & {\cellcolor[rgb]{0.922,0.902,0.902}}\textbf{0.511}    & {\cellcolor[rgb]{0.718,0.718,0.718}}0.101         & {\cellcolor[rgb]{0.922,0.902,0.902}}\textbf{0.086}   & {\cellcolor[rgb]{0.922,0.902,0.902}}\textbf{0.086}     \\ 
\hhline{|~~-------------|}
                                &                                    & \textbf{RTS}      & {\cellcolor[rgb]{0.718,0.718,0.718}}0.104         & \textbf{0.104}   & \textbf{0.103}    & {\cellcolor[rgb]{0.718,0.718,0.718}}0.262         & \textbf{0.251}   & \textbf{0.252}    & {\cellcolor[rgb]{0.718,0.718,0.718}}0.567         & \textbf{0.558}   & \textbf{0.551}    & {\cellcolor[rgb]{0.718,0.718,0.718}}0.110         & \textbf{0.110}   & \textbf{0.109}     \\ 
\hline
\multirow{8}{*}{\textbf{C\&W}}  & \multirow{4}{*}{\textbf{Resnet}}   & {\cellcolor[rgb]{0.922,0.902,0.902}}\textbf{Grad}     & {\cellcolor[rgb]{0.718,0.718,0.718}}0.042         & {\cellcolor[rgb]{0.922,0.902,0.902}}\textbf{0.041}   & {\cellcolor[rgb]{0.922,0.902,0.902}}\textbf{0.041}    & {\cellcolor[rgb]{0.718,0.718,0.718}}0.244         & {\cellcolor[rgb]{0.922,0.902,0.902}}\textbf{0.240}   & {\cellcolor[rgb]{0.922,0.902,0.902}}\textbf{0.242}    & {\cellcolor[rgb]{0.718,0.718,0.718}}0.546         & {\cellcolor[rgb]{0.922,0.902,0.902}}\textbf{0.545}   & {\cellcolor[rgb]{0.922,0.902,0.902}}\textbf{0.542}    & {\cellcolor[rgb]{0.718,0.718,0.718}}0.069         & {\cellcolor[rgb]{0.922,0.902,0.902}}\textbf{0.069}   & {\cellcolor[rgb]{0.922,0.902,0.902}}\textbf{0.069}     \\ 
\hhline{|~~-------------|}
                                &                                    & \textbf{CAM}      & {\cellcolor[rgb]{0.718,0.718,0.718}}0.050         & 0.060            & \textbf{0.049}    & {\cellcolor[rgb]{0.718,0.718,0.718}}0.328         & \textbf{0.307}   & 0.329             & {\cellcolor[rgb]{0.718,0.718,0.718}}0.573         & \textbf{0.570}   & 0.574             & {\cellcolor[rgb]{0.718,0.718,0.718}}0.038         & \textbf{0.037}   & \textbf{0.038}     \\ 
\hhline{|~~-------------|}
                                &                                    & {\cellcolor[rgb]{0.922,0.902,0.902}}\textbf{MASK}     & {\cellcolor[rgb]{0.718,0.718,0.718}}0.091         & {\cellcolor[rgb]{0.922,0.902,0.902}}\textbf{0.089}   & {\cellcolor[rgb]{0.922,0.902,0.902}}\textbf{0.089}    & {\cellcolor[rgb]{0.718,0.718,0.718}}0.221         & {\cellcolor[rgb]{0.922,0.902,0.902}}\textbf{0.220}   & {\cellcolor[rgb]{0.922,0.902,0.902}}\textbf{0.219}    & {\cellcolor[rgb]{0.718,0.718,0.718}}0.473         & {\cellcolor[rgb]{0.922,0.902,0.902}}0.474            & {\cellcolor[rgb]{0.922,0.902,0.902}}0.475             & {\cellcolor[rgb]{0.718,0.718,0.718}}0.081         & {\cellcolor[rgb]{0.922,0.902,0.902}}\textbf{0.081}   & {\cellcolor[rgb]{0.922,0.902,0.902}}\textbf{0.081}     \\ 
\hhline{|~~-------------|}
                                &                                    & \textbf{RTS}      & {\cellcolor[rgb]{0.718,0.718,0.718}}0.092         & \textbf{0.091}   & \textbf{0.092}    & {\cellcolor[rgb]{0.718,0.718,0.718}}0.308         & 0.310            & 0.311             & {\cellcolor[rgb]{0.718,0.718,0.718}}0.639         & \textbf{0.638}   & \textbf{0.638}    & {\cellcolor[rgb]{0.718,0.718,0.718}}0.086         & \textbf{0.085}   & \textbf{0.085}     \\ 
\hhline{|~--------------|}
                                & \multirow{4}{*}{\textbf{Densenet}} & {\cellcolor[rgb]{0.922,0.902,0.902}}\textbf{Grad}     & {\cellcolor[rgb]{0.718,0.718,0.718}}0.045         & {\cellcolor[rgb]{0.922,0.902,0.902}}\textbf{0.044}   & {\cellcolor[rgb]{0.922,0.902,0.902}}\textbf{0.044}    & {\cellcolor[rgb]{0.718,0.718,0.718}}0.241         & {\cellcolor[rgb]{0.922,0.902,0.902}}\textbf{0.241}   & {\cellcolor[rgb]{0.922,0.902,0.902}}\textbf{0.237}    & {\cellcolor[rgb]{0.718,0.718,0.718}}0.562         & {\cellcolor[rgb]{0.922,0.902,0.902}}\textbf{0.562}   & {\cellcolor[rgb]{0.922,0.902,0.902}}\textbf{0.561}    & {\cellcolor[rgb]{0.718,0.718,0.718}}0.056         & {\cellcolor[rgb]{0.922,0.902,0.902}}\textbf{0.055}   & {\cellcolor[rgb]{0.922,0.902,0.902}}\textbf{0.054}     \\ 
\hhline{|~~-------------|}
                                &                                    & \textbf{CAM}      & {\cellcolor[rgb]{0.718,0.718,0.718}}0.065         & 0.070            & \textbf{0.065}    & {\cellcolor[rgb]{0.718,0.718,0.718}}0.335         & \textbf{0.333}   & \textbf{0.335}    & {\cellcolor[rgb]{0.718,0.718,0.718}}0.570         & 0.572            & 0.571             & {\cellcolor[rgb]{0.718,0.718,0.718}}0.035         & 0.040            & \textbf{0.035}     \\ 
\hhline{|~~-------------|}
                                &                                    & {\cellcolor[rgb]{0.922,0.902,0.902}}\textbf{MASK}     & {\cellcolor[rgb]{0.718,0.718,0.718}}0.092         & {\cellcolor[rgb]{0.922,0.902,0.902}}\textbf{0.092}   & {\cellcolor[rgb]{0.922,0.902,0.902}}0.093             & {\cellcolor[rgb]{0.718,0.718,0.718}}0.220         & {\cellcolor[rgb]{0.922,0.902,0.902}}\textbf{0.220}   & {\cellcolor[rgb]{0.922,0.902,0.902}}\textbf{0.218}    & {\cellcolor[rgb]{0.718,0.718,0.718}}0.479         & {\cellcolor[rgb]{0.922,0.902,0.902}}0.481            & {\cellcolor[rgb]{0.922,0.902,0.902}}\textbf{0.476}    & {\cellcolor[rgb]{0.718,0.718,0.718}}0.080         & {\cellcolor[rgb]{0.922,0.902,0.902}}\textbf{0.079}   & {\cellcolor[rgb]{0.922,0.902,0.902}}\textbf{0.079}     \\ 
\hhline{|~~-------------|}
                                &                                    & \textbf{RTS}      & {\cellcolor[rgb]{0.718,0.718,0.718}}0.095         & \textbf{0.094}   & \textbf{0.094}    & {\cellcolor[rgb]{0.718,0.718,0.718}}0.289         & \textbf{0.286}   & 0.291             & {\cellcolor[rgb]{0.718,0.718,0.718}}0.558         & \textbf{0.558}   & 0.559             & {\cellcolor[rgb]{0.718,0.718,0.718}}0.092         & \textbf{0.091}   & \textbf{0.091}     \\ 
\hline
\multirow{8}{*}{\textbf{StAdv}} & \multirow{4}{*}{\textbf{Resnet}}   & {\cellcolor[rgb]{0.922,0.902,0.902}}\textbf{Grad}     & {\cellcolor[rgb]{0.718,0.718,0.718}}0.036         & {\cellcolor[rgb]{0.922,0.902,0.902}}\textbf{0.035}   & {\cellcolor[rgb]{0.922,0.902,0.902}}\textbf{0.035}    & {\cellcolor[rgb]{0.718,0.718,0.718}}0.259         & {\cellcolor[rgb]{0.922,0.902,0.902}}\textbf{0.229}   & {\cellcolor[rgb]{0.922,0.902,0.902}}\textbf{0.231}    & {\cellcolor[rgb]{0.718,0.718,0.718}}0.550         & {\cellcolor[rgb]{0.922,0.902,0.902}}\textbf{0.539}   & {\cellcolor[rgb]{0.922,0.902,0.902}}\textbf{0.540}    & {\cellcolor[rgb]{0.718,0.718,0.718}}0.032         & {\cellcolor[rgb]{0.922,0.902,0.902}}0.035            & {\cellcolor[rgb]{0.922,0.902,0.902}}0.035              \\ 
\hhline{|~~-------------|}
                                &                                    & \textbf{CAM}      & {\cellcolor[rgb]{0.718,0.718,0.718}}0.078         & \textbf{0.076}   & \textbf{0.077}    & {\cellcolor[rgb]{0.718,0.718,0.718}}0.141         & \textbf{0.114}   & \textbf{0.115}    & {\cellcolor[rgb]{0.718,0.718,0.718}}0.542         & \textbf{0.536}   & \textbf{0.537}    & {\cellcolor[rgb]{0.718,0.718,0.718}}0.023         & 0.026            & 0.026              \\ 
\hhline{|~~-------------|}
                                &                                    & {\cellcolor[rgb]{0.922,0.902,0.902}}\textbf{MASK}     & {\cellcolor[rgb]{0.718,0.718,0.718}}0.084         & {\cellcolor[rgb]{0.922,0.902,0.902}}0.086            & {\cellcolor[rgb]{0.922,0.902,0.902}}0.086             & {\cellcolor[rgb]{0.718,0.718,0.718}}0.246         & {\cellcolor[rgb]{0.922,0.902,0.902}}\textbf{0.243}   & {\cellcolor[rgb]{0.922,0.902,0.902}}\textbf{0.246}    & {\cellcolor[rgb]{0.718,0.718,0.718}}0.511         & {\cellcolor[rgb]{0.922,0.902,0.902}}\textbf{0.501}   & {\cellcolor[rgb]{0.922,0.902,0.902}}\textbf{0.502}    & {\cellcolor[rgb]{0.718,0.718,0.718}}0.074         & {\cellcolor[rgb]{0.922,0.902,0.902}}0.075            & {\cellcolor[rgb]{0.922,0.902,0.902}}0.076              \\ 
\hhline{|~~-------------|}
                                &                                    & \textbf{RTS}      & {\cellcolor[rgb]{0.718,0.718,0.718}}0.079         & \textbf{0.078}   & \textbf{0.078}    & {\cellcolor[rgb]{0.718,0.718,0.718}}0.255         & \textbf{0.248}   & \textbf{0.249}    & {\cellcolor[rgb]{0.718,0.718,0.718}}0.550         & \textbf{0.548}   & \textbf{0.547}    & {\cellcolor[rgb]{0.718,0.718,0.718}}0.022         & \textbf{0.022}   & \textbf{0.022}     \\ 
\hhline{|~--------------|}
                                & \multirow{4}{*}{\textbf{Densenet}} & {\cellcolor[rgb]{0.922,0.902,0.902}}\textbf{Grad}     & {\cellcolor[rgb]{0.718,0.718,0.718}}0.041         & {\cellcolor[rgb]{0.922,0.902,0.902}}\textbf{0.040}   & {\cellcolor[rgb]{0.922,0.902,0.902}}\textbf{0.040}    & {\cellcolor[rgb]{0.718,0.718,0.718}}0.250         & {\cellcolor[rgb]{0.922,0.902,0.902}}\textbf{0.221}   & {\cellcolor[rgb]{0.922,0.902,0.902}}\textbf{0.222}    & {\cellcolor[rgb]{0.718,0.718,0.718}}0.547         & {\cellcolor[rgb]{0.922,0.902,0.902}}\textbf{0.538}   & {\cellcolor[rgb]{0.922,0.902,0.902}}\textbf{0.539}    & {\cellcolor[rgb]{0.718,0.718,0.718}}0.034         & {\cellcolor[rgb]{0.922,0.902,0.902}}0.036            & {\cellcolor[rgb]{0.922,0.902,0.902}}0.036              \\ 
\hhline{|~~-------------|}
                                &                                    & \textbf{CAM}      & {\cellcolor[rgb]{0.718,0.718,0.718}}0.088         & \textbf{0.085}   & \textbf{0.085}    & {\cellcolor[rgb]{0.718,0.718,0.718}}0.172         & \textbf{0.151}   & \textbf{0.151}    & {\cellcolor[rgb]{0.718,0.718,0.718}}0.545         & \textbf{0.541}   & \textbf{0.541}    & {\cellcolor[rgb]{0.718,0.718,0.718}}0.062         & 0.064            & \textbf{0.062}     \\ 
\hhline{|~~-------------|}
                                &                                    & {\cellcolor[rgb]{0.922,0.902,0.902}}\textbf{MASK}     & {\cellcolor[rgb]{0.718,0.718,0.718}}0.085         & {\cellcolor[rgb]{0.922,0.902,0.902}}\textbf{0.085}   & {\cellcolor[rgb]{0.922,0.902,0.902}}\textbf{0.084}    & {\cellcolor[rgb]{0.718,0.718,0.718}}0.229         & {\cellcolor[rgb]{0.922,0.902,0.902}}\textbf{0.216}   & {\cellcolor[rgb]{0.922,0.902,0.902}}\textbf{0.219}    & {\cellcolor[rgb]{0.718,0.718,0.718}}0.493         & {\cellcolor[rgb]{0.922,0.902,0.902}}\textbf{0.485}   & {\cellcolor[rgb]{0.922,0.902,0.902}}\textbf{0.489}    & {\cellcolor[rgb]{0.718,0.718,0.718}}0.067         & {\cellcolor[rgb]{0.922,0.902,0.902}}\textbf{0.067}   & {\cellcolor[rgb]{0.922,0.902,0.902}}0.068              \\ 
\hhline{|~~-------------|}
                                &                                    & \textbf{RTS}      & {\cellcolor[rgb]{0.718,0.718,0.718}}0.082         & \textbf{0.081}   & \textbf{0.081}    & {\cellcolor[rgb]{0.718,0.718,0.718}}0.241         & \textbf{0.235}   & \textbf{0.235}    & {\cellcolor[rgb]{0.718,0.718,0.718}}0.543         & \textbf{0.539}   & \textbf{0.540}    & {\cellcolor[rgb]{0.718,0.718,0.718}}0.028         & 0.029            & 0.029              \\
\hline
\end{tabular}}
\end{table*}

Additionally, we continue our observation quantitatively. Table \ref{tab:l1_transferability} shows the $\mathcal{L}_{1}$ distance between attribution maps of adversarial (ADV$^2$, \ours and \oursplus) and benign samples across different interpreters. Observe that our methods generate better-quality interpretation maps on different interpretation models $g$ as compared to the samples produced by ADV$^2$. Overall, \ours and \oursplus maintain high transferability across interpreters for most of classes. 

%

\BfPara{Limitations}
We note that some adversarial inputs crafted on a specific interpreter $g$ may not be transferable to another interpreter $g'$ as different interpreters utilize different properties of DNNs to generate attribution maps.
Figure \ref{fig:transferability_limitation} displays some cases where using interpretations from one interpreter to another produces attribution maps that are not entirely similar to benign cases. However, the attribution maps of adversarial inputs from different interpretation models are still focused on the object in the image. 
Figure \ref{fig:transferability_limitation} shows some of these cases across different attacks (ADV$^2$, \ours and \oursplus). In summary, \ours and \oursplus may not generate transferable adversarial input in some cases. However, the interpretation focus (highlighted area) is still in the object. Thus, the attribution map can be valid if the user does not have knowledge about the benign interpretation.


\begin{figure}[h]
    \centering
    \captionsetup{justification=justified}
    \includegraphics[width=0.45\textwidth]{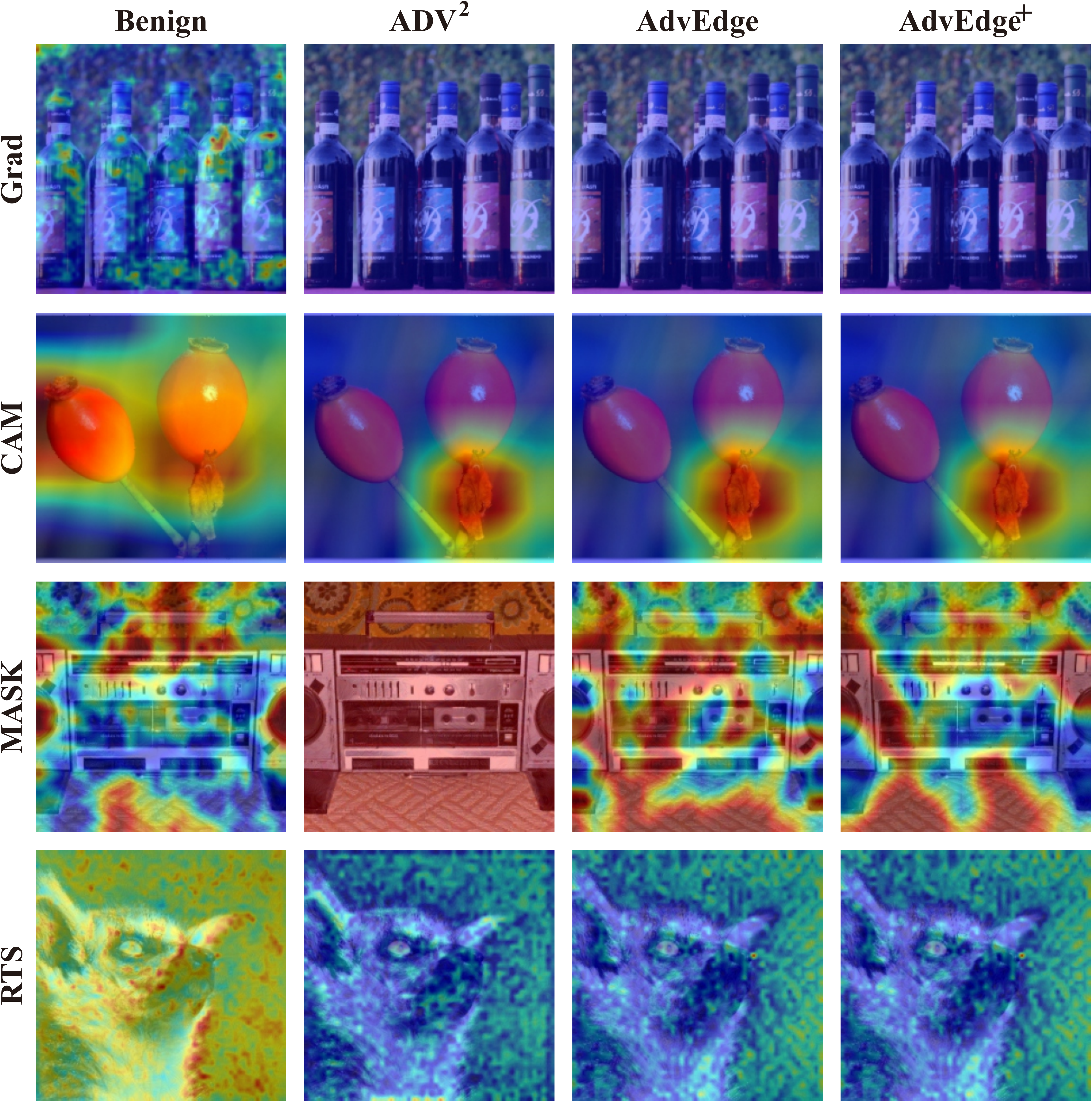}
    \caption{Outliers of adversarial inputs with interpretations similar to benign. Grad and MASK are based on StAdv and PGD, respectively (DenseNet-169). CAM and RTS interpreters are based on StAdv and C\&W, respectively (Resnet-50).}
    \label{fig:limit_discussion}
\end{figure}

\section{Discussion} \label{discussion}
As displayed in Figure \ref{fig:lp_all}, there are several cases that the attacks lack searching proper perturbation to mislead the target DNN and their coupled interpreters. We provide some examples that are considered as outliers for different attack frameworks on both ResNet and DenseNet in Figure \ref{fig:limit_discussion}. It is observed that as the interpretation (heatmap) is scattered across the whole image region (cases in MASK, Grad, and RTS), the attacks ADV$^2$, \ours and \oursplus cannot find the best adversarial space to meet the requirements of classifiers and interpreters. Additionally, samples that are located near decision boundaries are likely to jump into another class space. Adding small amount of perturbation causes the focus to be changed totally into another object in the image, attribution maps of which are totally different from the benign ones and it can be seen as the example of CAM interpreter. \ours and \oursplus may lack in finding the proper adversarial space in cases when the sample is near the decision boundary or the interpretation is spread over the image region.   


\section{Potential Countermeasures} \label{countermeasure}
In this section, we discuss the potential countermeasure against our proposed attack. Based on the results provided in the above sections, the adversarial interpretations of \ours and \oursplus share a high similarity with benign attribution maps. It is also mentioned that different interpretations utilize different behavioral aspects of DNN models. Using ensemble interpretation techniques can be encouraging to defend against interpretation-based attacks. However, aggregating an interpretation via the ensemble technique is quite challenging as the interpretations of different interpreters are contrasting. Additionally, the proposed attack can be adaptive by optimizing the interpretation loss with respect to all the interpreters; nevertheless, it is computationally expensive to generate an adversarial sample considering all the interpreters. Taking those into consideration, we recommend some countermeasures against the attack. 

\BfPara{Adversarial Detector} We suggest a multiple-interpreter-based detector that utilizes multiple interpretations of a single input to check whether the given input is adversarial or benign. Detectors can be efficacious at detecting adversarial samples via the usage of multiple interpreters to some extent.
For our experiment, we generated adversarial samples based on one interpreter and produce attribution maps of those samples via two interpreters including the target interpreter. For example, adversarial samples are generated based on Grad interpreter and attributions maps of those samples are extracted from both Grad and CAM interpreters. We repeated the same process by targeting CAM and applying Grad as a secondary interpreter. The generated attribution maps are based on single-channel, therefore, we stacked single-channel attribution maps of two interpreters to convert into benign and adversarial two-channel data correspondingly. 2,000 benign and 2,000 adversarial samples are produced for each experiment. As the dataset size is small, we adopted pre-trained DNN model VGG-16 \cite{simonyan2014very} to extract feature vectors from convolutional layers and we adopted ML model Gradient boosting classifier as a final layer instead of a fully-connected layer. The main reason for the approach is that the benign and adversarial attribution maps share high similarity and it is a complex process enough to classify the samples. Gradient boosting classifier is a group of machine learning algorithms that combines several weak learning models to form a strong predictive model.

\begin{table}[t]
\centering
\caption{The results of the two-interpretation-based detector. The upper part relates to 2-channel samples and the lower part shows the results of 3-channel samples.}
\label{tab:two_int_detector}
\arrayrulecolor{black}
\begin{tabular}{!{\color{black}\vrule}c!{\color{black}\vrule}c!{\color{black}\vrule}c!{\color{black}\vrule}c!{\color{black}\vrule}c!{\color{black}\vrule}} 
\hline
\multicolumn{5}{!{\color{black}\vrule}c!{\color{black}\vrule}}{\textbf{2-channel samples}}                                                                                                                                                                       \\ 
\hline
                                                  & {\cellcolor[rgb]{0.718,0.718,0.718}}\textbf{Grad} & {\cellcolor[rgb]{0.718,0.718,0.718}}\textbf{CAM} & {\cellcolor[rgb]{0.718,0.718,0.718}}\textbf{MASK} & {\cellcolor[rgb]{0.718,0.718,0.718}}\textbf{RTS}  \\ 
\hline
{\cellcolor[rgb]{0.718,0.718,0.718}}\textbf{Grad} & -                                                 & -                                                & -                                                 & -                                                 \\ 
\hline
{\cellcolor[rgb]{0.718,0.718,0.718}}\textbf{CAM}  & 0.896                                             & -                                                & -                                                 & -                                                 \\ 
\hline
{\cellcolor[rgb]{0.718,0.718,0.718}}\textbf{MASK} & 0.869                                             & 0.866                                            & -                                                 & -                                                 \\ 
\hline
{\cellcolor[rgb]{0.718,0.718,0.718}}\textbf{RTS}  & 0.762                                             & 0.716                                            & 0.701                                             & -                                                 \\ 
\hline
\multicolumn{5}{!{\color{black}\vrule}c!{\color{black}\vrule}}{\textbf{3-channel samples}}                                                                                                                                                                                \\ 
\hline
{\cellcolor[rgb]{0.718,0.718,0.718}}\textbf{Grad} & -                                                 & -                                                & -                                                 & -                                                 \\ 
\hline
{\cellcolor[rgb]{0.718,0.718,0.718}}\textbf{CAM}  & 0.890                                             & -                                                & -                                                 & -                                                 \\ 
\hline
{\cellcolor[rgb]{0.718,0.718,0.718}}\textbf{MASK} & 0.911                                             & 0.894                                            & -                                                 & -                                                 \\ 
\hline
{\cellcolor[rgb]{0.718,0.718,0.718}}\textbf{RTS}  & 0.744                                             & 0.712                                            & 0.714                                             & -                                                 \\
\hline
\end{tabular}
\arrayrulecolor{black}
\end{table}

As mentioned earlier, each attribution map is a one-channel image. As a result, we replaced the input and the output layer of VGG-16 and trained only those layers to adjust the weights of the model and to generate 2-channel samples. For the second case, we stacked the multiplication of attribution maps collected from two interpreters as the third channel. In Table \ref{tab:two_int_detector}, the results of both cases are provided. As shown Table \ref{tab:two_int_detector}, the detector could be used to improve the adversarial detection process by generating enough samples. We also created an experiment by adopting three different interpreters and stacked the attribution maps of those interpretation models into a 3-channel sample. Table \ref{tab:three_int_detector} shows results of the detector that is based on the combination of three interpreters.

\begin{table}[t]
\centering
\caption{Results of the three-interpretation-based detector.}
\label{tab:three_int_detector}
\scalebox{1.2}{
\begin{tabular}{|c|c|} 
\hline
\textbf{Combinations of interpreters}               & \textbf{Accuracy}  \\ 
\hline
{\cellcolor[rgb]{0.718,0.718,0.718}}Grad, CAM, MASK & 0.975              \\ 
\hline
{\cellcolor[rgb]{0.718,0.718,0.718}}Grad, CAM, RTS  & 0.861              \\ 
\hline
{\cellcolor[rgb]{0.718,0.718,0.718}}Grad, MASK, RTS & 0.838              \\ 
\hline
{\cellcolor[rgb]{0.718,0.718,0.718}}CAM, MASK, RTS  & 0.848              \\
\hline
\end{tabular}}
\end{table}

\BfPara{Interpretation-based Robust Training} The defense technique is based on the method in \cite{boopathy2020proper}. The technique considers a generic form of $L_1$ 2-class interpretation difference:
\begin{equation} \label{eq:int_adv_loss}
    \begin{aligned}
        \mathbb{D}_{L_1}(x, \hat{x}) = (1/2)(\|g_y(x, f) - g_y(\hat{x}, f)\|_1 \\ + \|g_{y_t}(x, f) - g_{y_t}(\hat{x}, f)\|_1)
    \end{aligned}
\end{equation}

In Eq. \ref{eq:int_adv_loss} creates a perturbation-independent lower bound for any adversarial attack that makes hard to fool a classifier and evade the interpretation discrepancy. Training a classifier against the worst-case interpretation difference is recommended by the min-max optimization problem:
\begin{equation} \label{eq:min_max_adv_train}
    \begin{aligned}
    \min_{\theta} \mathbb{E}_{x}\big[f_{train}(\theta; x, y) + \gamma \mathbb{D}_{worst}(x, \hat{x})\big]
    \end{aligned}
\end{equation}
where $f_{train}$ is the training loss (\ie cross-entropy loss), $\mathbb{D}_{worst}$ denotes a measurement of the highest interpretation difference between benign and adversarial samples $x$ and $\hat{x}$, and $\gamma$ balances the accuracy and model interpretability. For the experiment, we adopted PGD attack framework to generate adversarial samples. The results of the robust training are provided in Table \ref{tab:adv_train_result}. In the table, the results are produced by the Wide-ResNet \cite{zagoruyko2016wide}  model with coupled CAM interpreter on CIFAR-10 \cite{krizhevsky2009learning} dataset. Interpretability results are calculated by applying Kendall's Tau order rank correlation between the original and adversarial interpretation maps. As shown in the Table \ref{tab:adv_train_result}, the adversarial-trained model provided high classification and interpretation robustness as compared with the normal-trained model.

\begin{table}[t]
\centering
\caption{The results of interpretation-based robust training with different perturbation size $\epsilon$. NT stands for Normal Training and AT denotes Adversarial Training. Testing accuracy is based on the robustness of the classifier (upper part) and lower part provides the results of similarity of attribution maps.}
\label{tab:adv_train_result}
\begin{tabular}{|c|c|c|c|c|c|c|c|} 
\hline
\rowcolor[rgb]{0.718,0.718,0.718} \textbf{\textit{$\epsilon$}} & \textit{0}     & \textit{2/255} & \textit{4/255} & \textit{6/255} & \textit{8/255} & \textit{9/255} & \textit{10/255}  \\ 
\hline
\multicolumn{8}{|c|}{{\cellcolor[rgb]{0.718,0.718,0.718}}\textbf{Testing Accuracy}}                                                                                                            \\ 
\hline
\textbf{NT}                                         & \textbf{0.785} & 0.190          & 0.090          & 0.080          & 0.085          & 0.085          & 0.080            \\ 
\hline
\textbf{AT}                                    & 0.685          & \textbf{0.565} & \textbf{0.465} & \textbf{0.335} & \textbf{0.280} & \textbf{0.265} & \textbf{0.220}   \\ 
\hline
\multicolumn{8}{|c|}{{\cellcolor[rgb]{0.718,0.718,0.718}}\textbf{Attack against interpretability}}                                                                                             \\ 
\hline
\multicolumn{2}{|c|}{\textbf{NT}}                                    & 0.542          & 0.132          & -0.06          & -0.100         & -0.083         & -0.183           \\ 
\hline
\multicolumn{2}{|c|}{\textbf{AT}}                               & \textbf{0.870} & \textbf{0.782} & \textbf{0.736} & \textbf{0.703} & \textbf{0.706} & \textbf{0.676}   \\
\hline
\end{tabular}\vspace{-2em}
\end{table}

\section{Conclusion} \label{conclusion}
This paper proposes two attack methods (\ours and \oursplus) to improve the adversarial attacks on interpretable deep learning systems (IDLSes). Those methods use edge information of image inputs to optimize the ADV$^2$ attack, which provides adversarial samples to mislead the target DNN models and their coupled interpretation models at the same time. Through empirical examination of a large dataset on two distinct DNN architectures by adopting three different attack frameworks (PGD, StAdv, and C\&W), we showed the validity and efficacy of \ours and \oursplus attacks.
We presented our findings in comparison to four distinct types of interpreters (\ie Grad, CAM, MASK, and RTS).
The results indicated that \ours and \oursplus effectively produce adversarial samples that are capable of misleading DNN models and their interpreters. We also showed that the transferability of attribution maps generated by our attack methods provides significantly better results than existing methods. 
The results confirmed that the interpretability of IDLSes provides a limited sense of security in the decision-making process.

\section*{Acknowledgment}
This research was supported by the MSIT (Ministry of Science and ICT), Korea, under the ICT Creative Consilience Program (IITP-2021-2020-0-01821) supervised by the IITP (Institute for Information \& communications Technology Planning \& Evaluation), and the National Research Foundation of Korea (NRF) grant funded by the Korea government (MSIT) (No. 2021R1A2C1011198).

\balance
\bibliography{ref.bib}

\begin{thebibliography}{10}
\providecommand{\url}[1]{#1}
\csname url@samestyle\endcsname
\providecommand{\newblock}{\relax}
\providecommand{\bibinfo}[2]{#2}
\providecommand{\BIBentrySTDinterwordspacing}{\spaceskip=0pt\relax}
\providecommand{\BIBentryALTinterwordstretchfactor}{4}
\providecommand{\BIBentryALTinterwordspacing}{\spaceskip=\fontdimen2\font plus
\BIBentryALTinterwordstretchfactor\fontdimen3\font minus
  \fontdimen4\font\relax}
\providecommand{\BIBforeignlanguage}[2]{{%
\expandafter\ifx\csname l@#1\endcsname\relax
\typeout{** WARNING: IEEEtran.bst: No hyphenation pattern has been}%
\typeout{** loaded for the language `#1'. Using the pattern for}%
\typeout{** the default language instead.}%
\else
\language=\csname l@#1\endcsname
\fi
#2}}
\providecommand{\BIBdecl}{\relax}
\BIBdecl

\bibitem{Eldor}
E.~Abdukhamidov, M.~Abuhamad, F.~Juraev, E.~Chan-Tin, and T.~AbuHmed,
  ``Advedge: Optimizing adversarial perturbations against interpretable deep
  learning,'' in \emph{Computational Data and Social Networks}, D.~Mohaisen and
  R.~Jin, Eds.\hskip 1em plus 0.5em minus 0.4em\relax Cham: Springer
  International Publishing, 2021, pp. 93--105.

\bibitem{he2016deep}
K.~He, X.~Zhang, S.~Ren, and J.~Sun, ``Deep residual learning for image
  recognition,'' in \emph{Proceedings of the IEEE conference on computer vision
  and pattern recognition}, 2016, pp. 770--778.

\bibitem{sutskever2014sequence}
I.~Sutskever, O.~Vinyals, and Q.~V. Le, ``Sequence to sequence learning with
  neural networks,'' \emph{arXiv preprint arXiv:1409.3215}, 2014.

\bibitem{sabour2017dynamic}
S.~Sabour, N.~Frosst, and G.~E. Hinton, ``Dynamic routing between capsules,''
  \emph{arXiv preprint arXiv:1710.09829}, 2017.

\bibitem{zhang2018interpretable}
Q.~Zhang, Y.~N. Wu, and S.-C. Zhu, ``Interpretable convolutional neural
  networks,'' in \emph{Proceedings of the IEEE Conference on Computer Vision
  and Pattern Recognition}, 2018, pp. 8827--8836.

\bibitem{dabkowski2017real}
P.~Dabkowski and Y.~Gal, ``Real time image saliency for black box
  classifiers,'' \emph{arXiv preprint arXiv:1705.07857}, 2017.

\bibitem{fong2017interpretable}
R.~C. Fong and A.~Vedaldi, ``Interpretable explanations of black boxes by
  meaningful perturbation,'' in \emph{Proceedings of the IEEE International
  Conference on Computer Vision}, 2017, pp. 3429--3437.

\bibitem{simonyan2013deep}
K.~Simonyan, A.~Vedaldi, and A.~Zisserman, ``Deep inside convolutional
  networks: Visualising image classification models and saliency maps,''
  \emph{arXiv preprint arXiv:1312.6034}, 2013.

\bibitem{zhang2020interpretable}
X.~Zhang, N.~Wang, H.~Shen, S.~Ji, X.~Luo, and T.~Wang, ``Interpretable deep
  learning under fire,'' in \emph{29th $\{$USENIX$\}$ Security Symposium
  ($\{$USENIX$\}$ Security 20)}, 2020.

\bibitem{biggio2012poisoning}
B.~Biggio, B.~Nelson, and P.~Laskov, ``Poisoning attacks against support vector
  machines,'' \emph{arXiv preprint arXiv:1206.6389}, 2012.

\bibitem{dalvi2004adversarial}
N.~Dalvi, P.~Domingos, S.~Sanghai, and D.~Verma, ``Adversarial
  classification,'' in \emph{Proceedings of the tenth ACM SIGKDD international
  conference on Knowledge discovery and data mining}, 2004, pp. 99--108.

\bibitem{lecun2015deep}
Y.~LeCun, Y.~Bengio, and G.~Hinton, ``Deep learning,'' \emph{nature}, vol. 521,
  no. 7553, pp. 436--444, 2015.

\bibitem{juraev2022depth}
F.~Juraev, E.~Abdukhamidov, M.~Abuhamad, and T.~Abuhmed, ``Depth, breadth, and
  complexity: Ways to attack and defend deep learning models,'' in
  \emph{Proceedings of the 2022 ACM on Asia Conference on Computer and
  Communications Security}, 2022, pp. 1207--1209.

\bibitem{madry2017towards}
A.~Madry, A.~Makelov, L.~Schmidt, D.~Tsipras, and A.~Vladu, ``Towards deep
  learning models resistant to adversarial attacks,'' \emph{arXiv preprint
  arXiv:1706.06083}, 2017.

\bibitem{szegedy2013intriguing}
C.~Szegedy, W.~Zaremba, I.~Sutskever, J.~Bruna, D.~Erhan, I.~Goodfellow, and
  R.~Fergus, ``Intriguing properties of neural networks,'' \emph{arXiv preprint
  arXiv:1312.6199}, 2013.

\bibitem{abdukhamidov2022black}
E.~Abdukhamidov, F.~Juraev, M.~Abuhamad, and T.~Abuhmed, ``Black-box and
  target-specific attack against interpretable deep learning systems,'' in
  \emph{Proceedings of the 2022 ACM on Asia Conference on Computer and
  Communications Security}, 2022, pp. 1216--1218.

\bibitem{liu2016delving}
Y.~Liu, X.~Chen, C.~Liu, and D.~Song, ``Delving into transferable adversarial
  examples and black-box attacks,'' \emph{arXiv preprint arXiv:1611.02770},
  2016.

\bibitem{papernot2016transferability}
N.~Papernot, P.~McDaniel, and I.~Goodfellow, ``Transferability in machine
  learning: from phenomena to black-box attacks using adversarial samples,''
  \emph{arXiv preprint arXiv:1605.07277}, 2016.

\bibitem{huang2019enhancing}
Q.~Huang, I.~Katsman, H.~He, Z.~Gu, S.~Belongie, and S.-N. Lim, ``Enhancing
  adversarial example transferability with an intermediate level attack,'' in
  \emph{Proceedings of the IEEE/CVF international conference on computer
  vision}, 2019, pp. 4733--4742.

\bibitem{xie2019improving}
C.~Xie, Z.~Zhang, Y.~Zhou, S.~Bai, J.~Wang, Z.~Ren, and A.~L. Yuille,
  ``Improving transferability of adversarial examples with input diversity,''
  in \emph{Proceedings of the IEEE/CVF Conference on Computer Vision and
  Pattern Recognition}, 2019, pp. 2730--2739.

\bibitem{springenberg2014striving}
J.~T. Springenberg, A.~Dosovitskiy, T.~Brox, and M.~Riedmiller, ``Striving for
  simplicity: The all convolutional net,'' \emph{arXiv preprint
  arXiv:1412.6806}, 2014.

\bibitem{selvaraju2017grad}
R.~R. Selvaraju, M.~Cogswell, A.~Das, R.~Vedantam, D.~Parikh, and D.~Batra,
  ``Grad-cam: Visual explanations from deep networks via gradient-based
  localization,'' in \emph{Proceedings of the IEEE international conference on
  computer vision}, 2017, pp. 618--626.

\bibitem{zhou2016learning}
B.~Zhou, A.~Khosla, A.~Lapedriza, A.~Oliva, and A.~Torralba, ``Learning deep
  features for discriminative localization,'' in \emph{Proceedings of the IEEE
  conference on computer vision and pattern recognition}, 2016, pp. 2921--2929.

\bibitem{nguyen2015deep}
A.~Nguyen, J.~Yosinski, and J.~Clune, ``Deep neural networks are easily fooled:
  High confidence predictions for unrecognizable images,'' in \emph{Proceedings
  of the IEEE conference on computer vision and pattern recognition}, 2015, pp.
  427--436.

\bibitem{liu2018adversarial}
N.~Liu, H.~Yang, and X.~Hu, ``Adversarial detection with model
  interpretation,'' in \emph{Proceedings of the 24th ACM SIGKDD International
  Conference on Knowledge Discovery \& Data Mining}, 2018, pp. 1803--1811.

\bibitem{tao2018attacks}
G.~Tao, S.~Ma, Y.~Liu, and X.~Zhang, ``Attacks meet interpretability:
  Attribute-steered detection of adversarial samples,'' \emph{arXiv preprint
  arXiv:1810.11580}, 2018.

\bibitem{simonyan2014deep}
K.~Simonyan, A.~Vedaldi, and A.~Zisserman, ``Deep inside convolutional
  networks: Visualising image classification models and saliency maps,'' 2014.

\bibitem{moosavi2017universal}
S.-M. Moosavi-Dezfooli, A.~Fawzi, O.~Fawzi, and P.~Frossard, ``Universal
  adversarial perturbations,'' in \emph{Proceedings of the IEEE conference on
  computer vision and pattern recognition}, 2017, pp. 1765--1773.

\bibitem{kurakin2016adversarial}
A.~Kurakin, I.~Goodfellow, S.~Bengio \emph{et~al.}, ``Adversarial examples in
  the physical world,'' 2016.

\bibitem{carlini2017towards}
N.~Carlini and D.~Wagner, ``Towards evaluating the robustness of neural
  networks,'' in \emph{2017 ieee symposium on security and privacy (sp)}.\hskip
  1em plus 0.5em minus 0.4em\relax IEEE, 2017, pp. 39--57.

\bibitem{xiao2018spatially}
C.~Xiao, J.-Y. Zhu, B.~Li, W.~He, M.~Liu, and D.~Song, ``Spatially transformed
  adversarial examples,'' \emph{arXiv preprint arXiv:1801.02612}, 2018.

\bibitem{sobelsobel}
I.~Sobel, R.~Duda, and P.~Hart, ``Sobel-feldman operator.''

\bibitem{bach2015pixel}
S.~Bach, A.~Binder, G.~Montavon, F.~Klauschen, K.-R. M{\"u}ller, and W.~Samek,
  ``On pixel-wise explanations for non-linear classifier decisions by
  layer-wise relevance propagation,'' \emph{PloS one}, vol.~10, no.~7, p.
  e0130140, 2015.

\bibitem{smilkov2017smoothgrad}
D.~Smilkov, N.~Thorat, B.~Kim, F.~Vi{\'e}gas, and M.~Wattenberg, ``Smoothgrad:
  removing noise by adding noise,'' \emph{arXiv preprint arXiv:1706.03825},
  2017.

\bibitem{ancona2017towards}
M.~Ancona, E.~Ceolini, C.~{\"O}ztireli, and M.~Gross, ``Towards better
  understanding of gradient-based attribution methods for deep neural
  networks,'' \emph{arXiv preprint arXiv:1711.06104}, 2017.

\bibitem{ronneberger2015u}
O.~Ronneberger, P.~Fischer, and T.~Brox, ``U-net: Convolutional networks for
  biomedical image segmentation,'' in \emph{International Conference on Medical
  image computing and computer-assisted intervention}.\hskip 1em plus 0.5em
  minus 0.4em\relax Springer, 2015, pp. 234--241.

\bibitem{recht2019imagenet}
B.~Recht, R.~Roelofs, L.~Schmidt, and V.~Shankar, ``Do imagenet classifiers
  generalize to imagenet?'' in \emph{International Conference on Machine
  Learning}.\hskip 1em plus 0.5em minus 0.4em\relax PMLR, 2019, pp. 5389--5400.

\bibitem{huang2017densely}
G.~Huang, Z.~Liu, L.~Van Der~Maaten, and K.~Q. Weinberger, ``Densely connected
  convolutional networks,'' in \emph{Proceedings of the IEEE conference on
  computer vision and pattern recognition}, 2017, pp. 4700--4708.

\bibitem{wang2004image}
Z.~Wang, A.~C. Bovik, H.~R. Sheikh, and E.~P. Simoncelli, ``Image quality
  assessment: from error visibility to structural similarity,'' \emph{IEEE
  transactions on image processing}, vol.~13, no.~4, pp. 600--612, 2004.

\bibitem{simonyan2014very}
K.~Simonyan and A.~Zisserman, ``Very deep convolutional networks for
  large-scale image recognition,'' \emph{arXiv preprint arXiv:1409.1556}, 2014.

\bibitem{boopathy2020proper}
A.~Boopathy, S.~Liu, G.~Zhang, C.~Liu, P.-Y. Chen, S.~Chang, and L.~Daniel,
  ``Proper network interpretability helps adversarial robustness in
  classification,'' in \emph{International Conference on Machine
  Learning}.\hskip 1em plus 0.5em minus 0.4em\relax PMLR, 2020, pp. 1014--1023.

\bibitem{zagoruyko2016wide}
S.~Zagoruyko and N.~Komodakis, ``Wide residual networks,'' \emph{arXiv preprint
  arXiv:1605.07146}, 2016.

\bibitem{krizhevsky2009learning}
A.~Krizhevsky, G.~Hinton \emph{et~al.}, ``Learning multiple layers of features
  from tiny images,'' 2009.

\end{thebibliography}
\bibliographystyle{IEEEtran}
\end{document}